\documentclass[12pt]{article}
\usepackage[dvips]{graphicx}
\usepackage{float}
\usepackage{epsfig}
\usepackage{ulem}
\usepackage{latexsym,amsmath,amsfonts,amssymb}
\usepackage[latin1]{inputenc}
\usepackage{rotating}
\usepackage[american]{babel}
\usepackage[dvips]{graphicx}
\usepackage{bbm}
\usepackage{color}
\usepackage{slashed}
\usepackage[unicode]{hyperref}
\usepackage{lscape}
\usepackage{bigints}
\usepackage{enumerate}
\usepackage[shortlabels]{enumitem}
\usepackage{tikz}
\usetikzlibrary{decorations.pathreplacing}
\usetikzlibrary{shapes}
\def\im{Invent. Math.}

\def\hat{\widehat}
\def\a{\alpha}
\def\b{\beta}
\def\c{\gamma}
\def\d{\delta}
\def\f{\phi}               
\def\vf{\varphi}  
\def\tvf{\tilde{\varphi}}
\def\vp{\varphi}
\def\g{\gamma}
\def\h{\eta}
\def\j{\psi}
\def\k{\kappa}                    
\def\l{\lambda}
\def\m{\mu}
\def\n{\nu}
\def\o{\omega}  \def\w{\omega}

\def\q{\theta}  \def\th{\theta}                  
\def\r{\rho}                                     
\def\s{\sigma}                                   
\def\t{\tau}
\def\u{\upsilon}
\def\x{\xi}
\def\z{\zeta}
\def\pt{\tilde{\varphi}}
\def\tt{\tilde{\theta}}
\def\lab{\label}
\def\6{\partial}
\def\wg{\wedge}
\def\bpsi{\bar{\psi}}
\def\bt{\bar{\theta}}
\def\bvf{\bar{\varphi}}

\DeclareMathOperator{\tr}{tr}

\newcommand{\be}{\begin{equation}}
\newcommand{\ee}{\end{equation}}
\newcommand{\beq}{\begin{equation}}
\newcommand{\eeq}{\end{equation}}
\newcommand{\bea}{\begin{eqnarray}}
\newcommand{\eea}{\end{eqnarray}}

\newcommand{\ba}{\begin{eqnarray}}
\newcommand{\ea}{\end{eqnarray}}

\newcommand{\beqs}{\begin{eqnarray}}
\newcommand{\eeqs}{\end{eqnarray}}
\newcommand{\bal}{\begin{aligned}}
\newcommand{\eal}{\end{aligned}}
\makeatletter
\newcommand\setItemnumber[1]{\setcounter{enum\romannumeral\@enumdepth}{\numexpr#1-1\relax}}
\makeatother
\begin{document}
\baselineskip=15.5pt
\pagestyle{plain}
\setcounter{page}{1}

\def\del{{\partial}}
\def\vev#1{\left\langle #1 \right\rangle}
\def\cn{{\cal N}}
\def\co{{\cal O}}


\def\IC{{\mathbb C}}
\def\IR{{\mathbb R}}
\def\IZ{{\mathbb Z}}
\def\RP{{\bf RP}}
\def\CP{{\bf CP}}
\def\Poincaré{{Poincar\'e }}
\def\tr{{\rm tr}}
\def\tp{{\tilde \Phi}}

\def\TL{\hfil$\displaystyle{##}$}
\def\TR{$\displaystyle{{}##}$\hfil}
\def\TC{\hfil$\displaystyle{##}$\hfil}
\def\TT{\hbox{##}}
\def\HLINE{\noalign{\vskip1\jot}\hline\noalign{\vskip1\jot}}
\def\seqalign#1#2{\vcenter{\openup1\jot
   \halign{\strut #1\cr #2 \cr}}}
\def\lbldef#1#2{\expandafter\gdef\csname #1\endcsname {#2}}
\def\eqn#1#2{\lbldef{#1}{(\ref{#1})}%
\begin{equation} #2 \label{#1} \end{equation}}
\def\eqalign#1{\vcenter{\openup1\jot
     \halign{\strut\span\TL & \span\TR\cr #1 \cr
    }}}

\def\eno#1{(\ref{#1})}
\def\href#1#2{#2}
\def\half{\frac{1}{2}}



\def\ads{{\it AdS}}
\def\adsp{{\it AdS}$_{p+2}$}
\def\cft{{\it CFT}}

\newcommand{\ber}{\begin{eqnarray}}
\newcommand{\eer}{\end{eqnarray}}

\newcommand{\beqar}{\begin{eqnarray}}
\newcommand{\cN}{{\cal N}}
\newcommand{\cO}{{\cal O}}
\newcommand{\cA}{{\cal A}}
\newcommand{\cT}{{\cal T}}
\newcommand{\cF}{{\cal F}}
\newcommand{\cC}{{\cal C}}
\newcommand{\cR}{{\cal R}}
\newcommand{\cW}{{\cal W}}
\newcommand{\eeqar}{\end{eqnarray}}
\newcommand{\tht}{\thteta}
\newcommand{\lm}{\lambda}\newcommand{\Lm}{\Lambda}


\newcommand{\nonu}{\nonumber}
\newcommand{\oh}{\displaystyle{\frac{1}{2}}}
\newcommand{\dsl}
   {\kern.06em\hbox{\raise.15ex\hbox{$/$}\kern-.56em\hbox{$\partial$}}}
\newcommand{\id}{i\!\!\not\!\partial}
\newcommand{\as}{\not\!\! A}
\newcommand{\ps}{\not\! p}
\newcommand{\ks}{\not\! k}
\newcommand{\D}{{\cal{D}}}
\newcommand{\dv}{d^2x}
\newcommand{\Z}{{\cal Z}}
\newcommand{\N}{{\cal N}}
\newcommand{\Dsl}{\not\!\! D}
\newcommand{\Bsl}{\not\!\! B}
\newcommand{\Psl}{\not\!\! P}

\newcommand{\eeqarr}{\end{eqnarray}}
\newcommand{\ZZ}{{\rm \kern 0.275em Z \kern -0.92em Z}\;}


\def\del{{\delta^{\hbox{\sevenrm B}}}} \def\ex{{\hbox{\rm e}}}
\def\azb{A_{\bar z}} \def\az{A_z} \def\bzb{B_{\bar z}} \def\bz{B_z}
\def\czb{C_{\bar z}} \def\cz{C_z} \def\dzb{D_{\bar z}} \def\dz{D_z}
\def\im{{\hbox{\rm Im}}} \def\mod{{\hbox{\rm mod}}} \def\tr{{\hbox{\rm Tr}}}
\def\ch{{\hbox{\rm ch}}} \def\imp{{\hbox{\sevenrm Im}}}
\def\trp{{\hbox{\sevenrm Tr}}} \def\vol{{\hbox{\rm Vol}}}
\def\rl{\Lambda_{\hbox{\sevenrm R}}} \def\wl{\Lambda_{\hbox{\sevenrm W}}}
\def\fc{{\cal F}_{k+\cox}} \def\vev{vacuum expectation value}
\def\nodiv{\mid{\hbox{\hskip-7.8pt/}}}
\def\ie{{\em i.e.}}
\def\ie{\hbox{\it i.e.}}

\def\CC{{\mathchoice
{\rm C\mkern-8mu\vrule height1.45ex depth-.05ex
width.05em\mkern9mu\kern-.05em}
{\rm C\mkern-8mu\vrule height1.45ex depth-.05ex
width.05em\mkern9mu\kern-.05em}
{\rm C\mkern-8mu\vrule height1ex depth-.07ex
width.035em\mkern9mu\kern-.035em}
{\rm C\mkern-8mu\vrule height.65ex depth-.1ex
width.025em\mkern8mu\kern-.025em}}}

\def\RR{{\rm I\kern-1.6pt {\rm R}}}
\def\NN{{\rm I\!N}}
\def\ZZ{{\rm Z}\kern-3.8pt {\rm Z} \kern2pt}
\def\IB{\relax{\rm I\kern-.18em B}}
\def\ID{\relax{\rm I\kern-.18em D}}
\def\II{\relax{\rm I\kern-.18em I}}
\def\IP{\relax{\rm I\kern-.18em P}}
\newcommand{\CS}{{\scriptstyle {\rm CS}}}
\newcommand{\CSs}{{\scriptscriptstyle {\rm CS}}}
\newcommand{\rc}{\nonumber\\}
\newcommand{\bear}{\begin{eqnarray}}
\newcommand{\eear}{\end{eqnarray}}

\newcommand{\LL}{{\cal L}}

\def\mani{{\cal M}}
\def\calo{{\cal O}}
\def\calb{{\cal B}}
\def\calw{{\cal W}}
\def\calz{{\cal Z}}
\def\cald{{\cal D}}
\def\calc{{\cal C}}

\def\to{\rightarrow}
\def\ele{{\hbox{\sevenrm L}}}
\def\ere{{\hbox{\sevenrm R}}}
\def\zb{{\bar z}}
\def\wb{{\bar w}}
\def\nodiv{\mid{\hbox{\hskip-7.8pt/}}}
\def\menos{\hbox{\hskip-2.9pt}}
\def\dr{\dot R_}
\def\drr{\dot r_}
\def\ds{\dot s_}
\def\da{\dot A_}
\def\dga{\dot \gamma_}
\def\ga{\gamma_}
\def\dal{\dot\alpha_}
\def\al{\alpha_}
\def\cl{{closed}}
\def\cls{{closing}}
\def\vev{vacuum expectation value}
\def\tr{{\rm Tr}}
\def\to{\rightarrow}
\def\too{\longrightarrow}


\def\a{\alpha}
\def\b{\beta}
\def\c{\gamma}
\def\d{\delta}
\def\e{\epsilon}           
\def\F{\Phi}
\def\f{\phi}               
\def\vf{\varphi}  \def\tvf{\tilde{\varphi}}
\def\vp{\varphi}
\def\g{\gamma}
\def\h{\eta}
\def\j{\psi}
\def\k{\kappa}                    
\def\l{\lambda}
\def\m{\mu}
\def\n{\nu}
\def\o{\omega}  \def\w{\omega}
\def\q{\theta}  \def\th{\theta}                  
\def\r{\rho}                                     
\def\s{\sigma}                                   
\def\t{\tau}
\def\u{\upsilon}
\def\x{\xi}
\def\X{\Xi}
\def\z{\zeta}
\def\pt{\tilde{\varphi}}
\def\tt{\tilde{\theta}}
\def\lab{\label}
\def\6{\partial}
\def\wg{\wedge}
\def\atanh{{\rm arctanh}}
\def\bpsi{\bar{\psi}}
\def\bt{\bar{\theta}}
\def\bvf{\bar{\varphi}}

%



\newfont{\namefont}{cmr10}
\newfont{\addfont}{cmti7 scaled 1440}
\newfont{\boldmathfont}{cmbx10}
\newfont{\headfontb}{cmbx10 scaled 1728}





\newcommand{\re}{\,\mathbb{R}\mbox{e}\,}
\newcommand{\hyph}[1]{$#1$\nobreakdash-\hspace{0pt}}
\providecommand{\abs}[1]{\lvert#1\rvert}
\newcommand{\Nugual}[1]{$\mathcal{N}= #1 $}
\newcommand{\sub}[2]{#1_\text{#2}}
\newcommand{\partfrac}[2]{\frac{\partial #1}{\partial #2}}
\newcommand{\bsp}[1]{\begin{equation} \begin{split} #1 \end{split} \end{equation}}
\newcommand{\calF}{\mathcal{F}}
\newcommand{\calO}{\mathcal{O}}
\newcommand{\calM}{\mathcal{M}}
\newcommand{\calV}{\mathcal{V}}
\newcommand{\bbZ}{\mathbb{Z}}
\newcommand{\bbC}{\mathbb{C}}
\newcommand{\cK}{{\cal K}}

\newcommand{\Thq}{\Theta\left(\r-\r_q\right)}
\newcommand{\Dq}{\d\left(\r-\r_q\right)}
\newcommand{\kten}{\kappa^2_{\left(10\right)}}
\newcommand{\pbi}[1]{\imath^*\left(#1\right)}
\newcommand{\ho}{\hat{\omega}}
\newcommand{\tth}{\tilde{\th}}
\newcommand{\tf}{\tilde{\f}}
\newcommand{\tj}{\tilde{\j}}
\newcommand{\tw}{\tilde{\omega}}
\newcommand{\tz}{\tilde{z}}
\newcommand{\prj}[2]{(\partial_r{#1})(\partial_{\j}{#2})-(\partial_r{#2})(\partial_{\j}{#1})}
\def\atanh{{\rm arctanh}}
\def\sech{{\rm sech}}
\def\csch{{\rm csch}}
\allowdisplaybreaks[1]

\def\red{\textcolor[rgb]{0.98,0.00,0.00}}

\newcommand{\Dan}[1] {{\textcolor{blue}{#1}}}

\numberwithin{equation}{section}

\newcommand{\Tr}{\mbox{Tr}}    


%

\setcounter{footnote}{0}
\renewcommand{\theequation}{{\rm\thesection.\arabic{equation}}}

\begin{titlepage}

\begin{center}

\vskip .5in 
\noindent

{\Large \bf{ Electrostatic Description of Five-dimensional SCFTs} }
\bigskip\medskip

Andrea Legramandi \footnote{andrea.legramandi@swansea.ac.uk} and Carlos Nunez\footnote{c.nunez@swansea.ac.uk}\\

\bigskip\medskip
{\small 
Department of Physics, Swansea University, Swansea SA2 8PP, United Kingdom}

\vskip .5cm 
\vskip .9cm 
     	{\bf Abstract }\vskip .1in
\end{center}

\noindent
In this paper we discuss an infinite class of AdS$_6$ backgrounds in Type IIB supergravity dual to five dimensional SCFTs whose low energy description is in terms of linear quiver theories. The quantisation of the Page charges imposes 
that each solution is determined once a convex, piece-wise linear function is specified. 
In the dual field theory, we interpret this function as encoding the ranks of colour and flavour groups in the associated quiver. We check our proposal with several examples and provide general expressions for the holographic central charge and the Wilson loop VEV. Some solutions outside this general class, with less clear quiver interpretation, are also discussed. 

\noindent
\vskip .5cm
\vskip .5cm
\vfill
\eject

\end{titlepage}

\setcounter{footnote}{0}

\small{
\tableofcontents}

\normalsize

\newpage
\renewcommand{\theequation}{{\rm\thesection.\arabic{equation}}}
\section{Introduction}

It is more than twenty years now, that the Maldacena conjecture \cite{Maldacena:1997re} motivates the study of both gravity and field theory  topics. One such lines of research is the study of supersymmetric and conformal field theories in diverse dimensions. 
In particular, efforts have been dedicated to  the classification of Type II or M-theory backgrounds with AdS$_{d+1}$ factors.
These backgrounds are proposed as holographic duals, encoding semi-classically the highly quantum dynamics of SCFTs in $d$ dimensions with different amounts of SUSY. For the case in which the solutions are half-maximal supersymmetric, important progress in classifying string backgrounds and the mapping to families of quantum field theories has been achieved. This is the framework in which  this work should be read.
Let us  summarise the field theory-string background correspondences, climbing-up in dimensions. 
\\
In the case of one dimensional conformal quantum mechanical theories, backgrounds with AdS$_2$ factors and  the associated quantum mechanical systems have been discussed. 
Half-maximal BPS backgrounds containing an AdS$_2$ factor were studied in \cite{Corbino:2018fwb}-\cite{Hong:2019wyi}. 
The recent works \cite{Dibitetto:2019nyz}-\cite{Lozano:2021rmk} made precise and concrete the correspondence, for different families of string backgrounds containing an AdS$_2$ factor.

The case of half-maximal two dimensional SCFTs was well studied. From the perspective advertised above (backgrounds and dual field theories) we encounter the works  \cite{Lozano:2015bra}-\cite{Filippas:2020qku}. For three dimensional ${\cal N}=4$ SCFTs, the field theoretical aspects presented in \cite{Gaiotto:2008ak} were discussed holographically in \cite{DHoker:2007hhe}-\cite{Lozano:2016wrs}, among other works.

 In the case of ${\cal N}=2$ SCFTs in four dimensions, the field theories studied in \cite{Gaiotto:2009we} have holographic duals first discussed in \cite{Gaiotto:2009gz}, and further elaborated (among other works) in \cite{ReidEdwards:2010qs}-\cite{Bah:2019jts}. Let us jump off the five-dimensional case (that will occupy the rest of this work), and state that
 an infinite family of six-dimensional ${\cal N}=(1,0)$ SCFTs was discussed  both from the field theoretical and holographic points of view in \cite{Apruzzi:2013yva}-\cite{Apruzzi:2015wna}, among other papers.

The case of ${\cal N}=1$ five dimensional SCFTs (with eight Poincar\'e supercharges) was analysed from the field theoretical and holographic viewpoints in many papers. Indeed, starting with the foundational work of Seiberg \cite{Seiberg:1996bd}, followed by \cite{Brandhuber:1999np}-\cite{Passias:2012vp}, to the more recent works by D'Hoker, Gutperle, Uhlemann and Karch \cite{DHoker:2016ujz}, a long list of papers testing  the correspondence and analysing predictions derived for this case, have been presented \cite{DHoker:2016ujz}-\cite{Uhlemann:2020bek}. The developments in five dimensional ${\cal N}=1$ SCFTs have a holographic side and a geometric engineering side, on which huge progress was also achieved, see for example \cite{DelZotto:2017pti}-\cite{Bhardwaj:2020gyu}. In this paper, as explained above, we focus on linking AdS$_6$ backgrounds in Type IIB supergravity with conformal field theories in five dimensions, along the lines of  \cite{DHoker:2016ujz}-\cite{Uhlemann:2020bek}. The backgrounds we find are slightly different from those in \cite{DHoker:2016ujz}-\cite{Uhlemann:2020bek}. When field theoretical observables are computed in the regime in which a comparison with supergravity is meaningful, we find the same results as those obtained in the picture developed in  \cite{DHoker:2016ujz}-\cite{Uhlemann:2020bek}. A possible advantage of the formalism presented here is that some other calculations and the interpretation of the solutions may be easier to perform using our 'electrostatic viewpoint'. For example, it is simple to add D7 branes in this language.

\subsection*{General idea of this paper}
The present work follows closely these lines of research. 
In fact, the construction of the half-SUSY holographic duals in all different dimensions mostly proceeds following this  'algorithm':
\begin{itemize}
\item{Write the most generic background dual to a $d$-dimensional SCFT. This contains an AdS$_{d+1}$ and (given the amount of SUSY we consider) at least an $SU(2)_R$ isometry, to reflect the R-symmetry of the eight Poincar\'e supercharges SCFT. Ramond and NS fields must be compatible with these isometries. The background metric reads,
\begin{equation}
ds^2= f_1 (\vec{y})AdS_{d+1} + f_2(\vec{y}) d\Omega_2 + f_3 (\vec{y})d\Sigma_{7-d}(\vec{y}).\nonumber
\end{equation}
The functions $f_i$ can only depend on the coordinates of $\Sigma_{7-d}(\vec{y})$. If the R-symmetry is bigger than $SU(2)$, this is realised geometrically and taken from the $(7-d)$ $\vec{y}$-coordinates inside $\Sigma_{7-d}$. If we work with eleven dimensional supergravity, the manifold $\Sigma$ has dimension $(8-d)$. The case that occupies us in this work has $d=5$. The internal coordinates are denoted as $\vec{y}=(\sigma,\eta)$.
}
\item{The BPS equations ensuring half-SUSY and Bianchi identity are imposed. These are a non-linear system of first and second order PDE's. Operating with these equations, sometimes one is able to write the $f_i$ and the fluxes in terms of a single function $V(\vec{y})$  and its derivatives where the function $V$ solves a {\it linear} PDE. Some situations are known for which the $V$-function solves a non-linear PDE \cite{Gaiotto:2009gz,Legramandi:2018itv,Imamura:2001cr}.}
\item{Reasonable boundary conditions need to be imposed. The quantisation of Page charges in the gravity background implies that one of these boundary conditions is written in terms of a 'Rank function', a convex polygonal, linear by pieces function, with integer values at integer points. }
\item{The Rank function is put in correspondence with a quiver field theory. The ranks of the gauge and flavour groups are encoded in this function. After the field theory-string background pair is identified, various tests and predictions of the correspondence follow.}
\end{itemize}
Up-to technical subtleties dependent on the dimension $d$ of the SCFT, this is how the construction of duals to half-maximal SCFTs in space-time dimension $d$ proceeds. There are two exceptions, for $d=3$ \cite{DHoker:2007hhe} and $d=5$ \cite{DHoker:2016ujz,DHoker:2016ysh}. In those cases, the authors solve the system in terms of holomorphic functions and their integrals, using the powerful machinery of complex analysis. In this paper, we follow the algorithm above to find duals to five dimensional ${\cal N}=1$ SCFTs.

Fortunately, the first two steps of the above `algorithm' have been covered in the papers \cite{Niall-Tomasiello},\cite{AGLMZ}.  We profit from these results and complete the other steps, proposing a precise relation between SCFTs and Type IIB AdS$_6$ background. In this way we are giving an alternative description to the one very well developed in \cite{DHoker:2016ujz}-\cite{Uhlemann:2020bek}. It is worth mentioning that Uhlemann gave the first steps in using the Rank function for the purpose of matrix model calculations  \cite{Uhlemann:2019ypp}, \cite{Uhlemann:2020bek}. 

The main reason for tackling a problem that was already extensively discussed is to develop a different language with which one can readily identify the dual field theory. For example, in this language it is easy to consider the effect of D7 branes and find many physically meaningful solutions. Moreover, having two descriptions of a system might help to think about some new problems and it may provide an alternative view of known results, suggesting a connection with other constructions. Following the steps outlined above, we start from the material of the paper \cite{AGLMZ}.  We  need to solve a linear PDE for the potential function $V(\sigma,\eta)$ with suitable boundary conditions. We discuss the correspondence between a five dimensional quiver field theory and a 'holographic electrostatic problem' and present checks of this correspondence.

The contents of this work are organised as follows. In Section \ref{sectiongeometry}, we discuss the Type IIB configuration. We find the solution to the linear PDE written in terms of a Fourier series.
The boundary conditions for the PDE are specified. Only for our choice, the Page charges of the background are quantised. Then, we analyse the behaviour of the solution at special points in the $(\sigma,\eta)$-plane. The presence of 'colour-branes' dissolved in fluxes and 'flavour branes' present in the background as a localised physical object (indicated by a violation of the Bianchi identities) is discussed. Hence our configuration is better thought in terms of a Type IIB solution plus branes, on which the global symmetries of the SCFT are realised.

In Section \ref{sec:QFT} we discuss some aspects of the QFT in five dimensions and the SCFTs they flow to in the UV. We draw a precise  correspondence between the SCFT, the Rank function used as boundary condition and the string background. We discuss some illustrative examples giving detailed expressions for the Rank function, the potential function $V(\sigma,\eta)$ and some observables such as the holographic central charge and the VEV of Wilson loops. These observables, that have been calculated using localisation in the IR QFT description, are exact result and hence provide a check of the correspondence proposed. We also derive generic expressions for the potential function and of these observables.

In Section \ref{section-special}, we discuss special backgrounds that do not strictly follow from the above mentioned algorithm. In particular, the reasonable boundary conditions are not satisfied. We explain that this afflicts the background with singularities and propose a field theoretical mechanism to cure the singular behaviour.

In Section \ref{concl} we summarise the results of this work, present some conclusions and propose some follow up themes of investigation. Generous appendices are written for the benefit of the reader wishing to work on these topics, comparing with the results of the papers   \cite{DHoker:2016ujz}-\cite{Uhlemann:2020bek}, presenting explicit derivations of the expressions along the paper and providing more explicit examples.

\section{Supergravity background}\label{sectiongeometry}
We first discuss  geometrical aspects of the supergravity solutions. We summarise the background preserving ${\cal N}=1$ SUSY if a linear PDE is satisfied. We solve the PDE, carefully analyse the singularity structure and quantised charges.
\subsection{The Type IIB Background}
In this  section we present a Type IIB background with an AdS$_6$ factor, preserving eight Poincar\'e supersymmetries. The solution contains a two sphere parameterized by some coordinate $(\theta,\varphi)$ and realising the $SU(2)_R$ symmetry of the dual SCFT; since the R-symmetry is preserved, all the fields and fluxes can depend on the $S^2$ just via its volume form Vol$(S^2)$. It was originally  found in  the paper \cite{AGLMZ}. We work with a background related to that in \cite{AGLMZ} by an S-duality as explained in Appendix \ref{appendixcomments}. The full configuration consist of a metric, dilaton, $B_2$-field in the NS sector and $C_2$ and $C_0$ in the Ramond sector. The configuration is written in terms of a potential function $V(\sigma,\eta)$ that solves a linear partial differential equation written below.
The type IIB background in string frame is,
\begin{eqnarray}
& & ds_{10,st}^2= f_1(\sigma,\eta)\Big[ds^2(\text{AdS}_6) + f_2(\sigma,\eta) d s^2 (S^2)+ f_3(\sigma,\eta)(d\sigma^2+d\eta^2) \Big],\;\;e^{-2\Phi}=f_6(\sigma,\eta) , \nonumber\\[2mm]
& & B_2=f_4(\sigma,\eta) \text{Vol}(S^2),\;\;C_2= f_5(\sigma,\eta) \text{Vol}(S^2),\;\;\; C_0= f_7(\sigma,\eta), \label{background}\\ [2mm]
& & f_1= \frac{2}{3}\sqrt{\sigma^2 +\frac{3\sigma \partial_\sigma V}{\partial^2_\eta V}},\;\; f_2= \frac{\partial_\sigma V \partial^2_\eta V}{3\Lambda},\;\;f_3= \frac{\partial^2_\eta V}{3\sigma \partial_\sigma V},\;\;\Lambda=\sigma(\partial_\sigma\partial_\eta V)^2 + (\partial_\sigma V-\sigma \partial^2_\sigma V)  \partial^2_\eta V,\nonumber\\[2mm]
& & f_4=\frac{2}{9}\left(\eta -\frac{(\sigma \partial_\sigma V) (\partial_\sigma\partial_\eta V)}{\Lambda} \right),\;\;\;\; f_5=4\left( V- \frac{\sigma\partial_\sigma V}{\Lambda} (\partial_\eta V (\partial_\sigma \partial_\eta V) -3 (\partial^2_\eta V)(\partial_\sigma V)) \right),\nonumber\\[2mm]
& & f_6=18^2 \frac{3\sigma^2 \partial_\sigma V \partial^2_\eta V}{(3 \partial_\sigma V +\sigma \partial^2_\eta V)^2}\Lambda,\;\;\;\; f_7=18\left( \partial_\eta V + \frac{(3\sigma \partial_\sigma V) (\partial_\sigma\partial_\eta V )}{3\partial_\sigma V +\sigma \partial^2_\eta V}  \right).\nonumber
\end{eqnarray}
The function $V(\sigma,\eta)$ solves
\begin{equation}
\partial_\sigma \left(\sigma^2 \partial_\sigma V\right) +\sigma^2 \partial^2_\eta V=0.\label{diffeq} \end{equation}
In what follows we study this PDE, proposing boundary conditions that lead to a  nice interpretation of the solutions. In Section \ref{pagesection}, we impose quantisation of the Page charges associated with the infinite family of solutions in eq.(\ref{background}) and avoid badly-singular behaviours.

\subsection{Resolution of the PDE}\label{resPDE}
Let us discuss solutions to the differential equation (\ref{diffeq}). It helps our intuition to make the change
\begin{equation}
V(\sigma,\eta)=\frac{\hat{V} (\sigma,\eta)}{\sigma},\label{change1}
\end{equation}
which implies that the PDE in (\ref{diffeq}) reads like a Laplace equation in flat space,
\begin{equation}
\partial^2_\sigma \hat{V} + \partial_\eta^2 \hat{V}=0.\label{eqfinal}
\end{equation}
We choose the variable $\eta$ to be bounded in the interval $[0,P]$ and $\sigma$ to range over the real axis $-\infty<\sigma<\infty$. We impose the boundary conditions,
\begin{eqnarray}
& & \hat{V}(\sigma\to\pm\infty,\eta)=0,\;\;\;\;\;\hat{V}(\sigma, \eta=0)= \hat{V}(\sigma, \eta=P)=0.\nonumber\\
& & \lim_{\epsilon\to 0}\left(\partial_\sigma \hat{V}(\sigma=+\epsilon,\eta)- \partial_\sigma \hat{V}(\sigma=-\epsilon,\eta)\right)= {\cal R}(\eta).\label{bc}
\end{eqnarray}

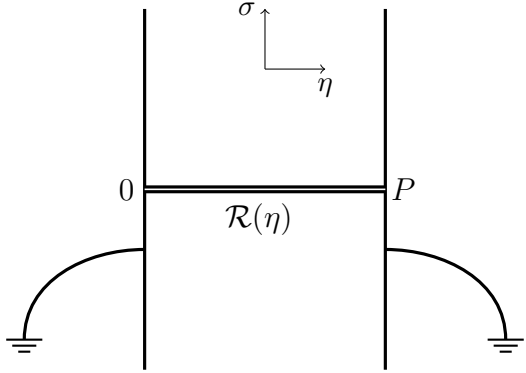
\begin{figure}
	\centering
	\begin{tikzpicture}[scale=0.8]
	\draw[very thick] (0,0) -- (0,6); 
	\draw[very thick] (4,0) -- (4,6);
	\draw[very thick] (0,2) to [out=-180,in=90] (-2,0.5);
	\draw[very thick] (4,2) to [out=0,in=90] (6,0.5);
	\draw[thick] (5.7,0.5) -- (6.3,0.5);
	\draw[thick] (5.8,0.4) -- (6.2,0.4);
	\draw[thick] (5.9,0.3) -- (6.1,0.3);
	\draw[thick] (-1.7,0.5) -- (-2.3,0.5);
	\draw[thick] (-1.8,0.4) -- (-2.2,0.4);
	\draw[thick] (-1.9,0.3) -- (-2.1,0.3);
	\draw[->] (2,5) -- (2,6);
	\draw[->] (2,5) -- (3,5);
	\node at (1.7,6) {\small $\sigma$};
	\node at (3,4.7) {\small $\eta$};
	\draw[very thick,double] (0,3) -- (4,3); 
	\node at (-0.3,3) {$0$};
	\node at (4.3,3) {$P$};
	\node at (1.9,2.5) {$\mathcal{R}(\eta)$};
	\end{tikzpicture}
	\caption{Depiction of the electrostatic problem for $\hat{V}$. The two conducting planes at $\eta=0,P$ have zero potential, while at $\sigma=0$ we have a charge distribution equal to $\cal R (\eta)$.}
	\label{fig:elect_problem}
\end{figure}

These can be interpreted as the boundary conditions for the electrostatic problem of two conducting planes (at zero electrostatic potential) as depicted in Figure \ref{fig:elect_problem}. The conducting planes extend over the the $\sigma$-direction and are placed at $\eta=0$ and $\eta=P$. We also have a charge density ${\cal R}(\eta)$ at $\sigma=0$, extended along $0\leq \eta\leq P$, as indicated by the difference of the normal components of the electric field. 
The function ${\cal R}(\eta)$ begins and  ends the $\eta$-direction, making it an interval,  according to 
\begin{equation}
{\cal R}(\eta=0)={\cal R} (\eta=P)=0.\label{condicion-f}\nonumber
\end{equation}
In Section \ref{pagesection} we discuss the conditions on ${\cal R}(\eta)$, as imposed by the quantisation of the Page charges.
We find a solution separating variables,
\begin{eqnarray}
& & \hat{V}(\sigma, \eta)=\Sigma(\sigma) E(\eta),\;\;\;\;\ddot{E}(\eta)+\lambda^2 E(\eta)=0,\;\;\;\; \Sigma''(\sigma) -\lambda^2 \Sigma(\sigma)=0. \label{solutions}
\end{eqnarray}
We solve in detail for the case $\lambda^2>0$. The cases $\lambda^2<0, \lambda=0$ do not satisfy the boundary conditions in eq.(\ref{bc}).
The solutions are
\begin{eqnarray}
& &E(\eta)= A \sin(\lambda \eta) + B \cos(\lambda \eta),\;\;\;\; \Sigma(\sigma)={C e^{-\lambda \sigma}} + {D e^{\lambda \sigma}}.\nonumber
\end{eqnarray}
Imposing the boundary conditions written in eq.(\ref{bc}) we find,
\begin{equation}
\label{eq:fourier_vhat} 
\hat{V}(\sigma,\eta)= \sum_{k=1}^\infty a_k \sin\left(\frac{k\pi}{P}\eta \right) {e^{-\frac{k\pi}{P}|\sigma|}},\;\;\;\;a_k= \frac{1}{\pi k }\int_{0}^P {\cal R}(\eta)\sin\left( \frac{k\pi}{P} \eta\right) ~d\eta.
\end{equation}
We have used that $\mathcal{R}$ can be expanded on a Fourier basis
\begin{equation}
{\cal R}(\eta)= \sum_{k=1}^\infty c_k \sin \left(\frac{k\pi}{P}\eta \right) ,\;\;\;\;\; 2\pi k a_k=-P c_k.\label{rankfunction}
\end{equation}

Notice that we can introduce a complex variable
\begin{equation}
z= \sigma - i\eta,\nonumber
\end{equation}
and write the potential $\hat{V}=\sigma V$ as a harmonic function for both $\sigma > 0$ and $\sigma < 0$
\begin{equation}
\hat{V}(\sigma,\eta)= \begin{cases}
\sum_{k=1}^\infty  \frac{a_k}{2 i} \left( e^{-\frac{k\pi}{P}z} - e^{-\frac{k \pi}{{P}} \bar{z} }\right) & \sigma \ge 0, \\[2mm]
\sum_{k=1}^\infty  \frac{i a_k}{2 } \left( e^{\frac{k\pi}{P} z} - e^{\frac{k \pi}{{P}} \bar{z} }\right) & \sigma < 0 .
\end{cases}\label{potenciales}
\end{equation}
$\hat{V}$ can therefore be expressed as the real part of a holomorphic function, and regularity is broken at $\sigma = 0$ due to the charge density in the electrostatic problem; see appendix \ref{sec:to_DGKU} for a more detailed discussion about how to translate our formalism in the holomorphic one in \cite{DHoker:2016ujz,DHoker:2016ysh}.

The reader can check that the potentials in eq.(\ref{potenciales}) solve the equations (\ref{diffeq}),(\ref{eqfinal}) subject to the conditions in eq.(\ref{bc}). We now move to analyse the behaviour of this class of solutions at special points.

\subsection{Behaviour at special points}

In this section we analyse the behaviour of the metric and the dilaton, as we approach special points. In particular we will focus on the points $\eta= 0, \eta= P$ and $\sigma \to \pm \infty$.

We start by considering the metric behaviour on the boundary at $\eta=0$. The discussion is identical for $\eta =P$. In this limit, we have that $f_1$ and $f_3$ are finite. Explicitly we have
\begin{align}
f_1^2(\sigma,0) =&  \frac49 \, \, \frac{\sum _{k=1}^\infty  \frac{\pi k}{P} a_k \left(\left(\frac{\pi  k \sigma }{P}\right)^2+3 | \sigma | \frac{\pi  k \sigma }{P} +3\right) e^{-\frac{\pi  k | \sigma | }{P}}}{\sum _{k=1}^\infty a_k \left(\frac{\pi  k}{P}\right)^3 e^{-\frac{\pi  k | \sigma | }{P}}} \, , \nonumber\\[2mm]
f_3(\sigma,0) =& \frac13 \, \, \frac{ \sum _{k=1}^\infty a_k \left(\frac{\pi  k}{P}\right)^3 e^{-\frac{\pi  k | \sigma | }{P}}}{ \sum _{k=1}^\infty a_k \left(\left(\frac{\pi  k}{P}\right)^2 | \sigma | + \frac{\pi k}{P} \right) e^{-\frac{\pi  k | \sigma | }{P}}} \, , \nonumber
\end{align}
while $f_2 \to \eta^2 f_3(\sigma,0)$. Using these expressions, we find that at these boundaries, the metric reads
\begin{equation}
\begin{split}
\eta \to 0 \, \qquad d s^2_{10} =& f_1(\sigma,0) \, \left(d s^2 (\text{AdS}_6) + f_3(\sigma,0) (\eta^2 d s^2 (S^2)+  d \eta^2 + d \sigma^2 ) \right) \, ,\\
\eta \to P \, \qquad d s^2_{10} =& f_1(\sigma,P) \left(d s^2 (\text{AdS}_6) + f_3(\sigma,P) ((\eta-P)^2 d s^2(S^2) +  d \eta^2 + d \sigma^2  ) \right) \, .
\end{split}
\end{equation}
In these two limits the metric is regular since it is given by a warped product AdS$_6 \times \mathbb{R}^4$ with warpings which are in general non-singular. The regularity of the solution at the boundary is confirmed by the regularity of the dilaton, which is finite. 

Let us  now consider the limit $\sigma \to \pm \infty$. We find that the leading contribution to  the potential $\hat{V}=\sigma V$ is given by the mode with $k=1$---see eq.(\ref{eq:fourier_vhat}). The asymptotic behaviours are,
\begin{equation}
\sigma V(\sigma,\eta) \sim  \partial^2_\eta (\sigma V) \sim \sin\left(\frac{\pi}{P}\eta \right)  e^{-\frac{\pi}{P}|\sigma|} \, , \quad \sigma^2 \partial_\sigma V \sim |\sigma| \sin\left(\frac{\pi}{P}\eta \right)  e^{-\frac{\pi}{P}|\sigma|} \, , \quad \Lambda \sim \sigma^{-1} e^{-\frac{2\pi}{P}|\sigma|} \, .
\end{equation}
Using these relations and up-to constant factors we have
\begin{equation}
\sigma \to \pm \infty \, \qquad d s^2 = |\sigma| d s^2 (\text{AdS}_6) + \sin^2 \left( \frac{\pi}{P} \eta\right) d s^2 (S^2)+  d \eta^2 + d \sigma^2 .
\end{equation}
A similar analysis for the dilaton leads to
\begin{equation}
\sigma \to \infty \qquad  e^{-\Phi} \sim \frac{ e^{- \frac{\pi}{P} |\sigma|}}{\sqrt{|\sigma|}} \, .
\end{equation}
Performing the change of coordinates $|\sigma| \to -\log r$ with $r$ small and positive, the metric and the dilaton display the behaviour of a $(p,q)$-five-brane for $\sigma\to\pm\infty$, as described in \cite{DHoker:2016ujz}.

The behaviour of this family of AdS$_6$ backgrounds at $\sigma = 0$ is characterized by some singularities in the fluxes. These can be studied by looking at the conserved Page charges. As we are going to discuss, the quantization conditions will lead to certain constraints on the possible Rank functions
${\cal R}(\eta)$.

\subsection{Page charges}\label{pagesection}
In this section, we study the conserved and quantised charges associated with the background in eq.(\ref{background}). We use the solutions described in Section \ref{resPDE}. As we find below, imposing quantisation on the charges restricts the function ${\cal R}(\eta)$ in eq.(\ref{eqfinal}) to be a convex piece-wise linear function.

In our conventions, the volume element on the sphere is defined by $\text{Vol}(S^2)= \sin\theta d\theta\wedge d\varphi$ and we write the field strengths
\begin{eqnarray}
& & H_3= dB_2= \left( \partial_\sigma f_4 d\sigma +\partial_\eta f_4 d\eta \right)\wedge \text{Vol}(S^2),\; \label{fieldstrength}\\[2mm]
& & \hat{F}_1= F_1=dC_0= \partial_\sigma f_7 d\sigma+\partial_\eta f_7 d\eta, \nonumber\\[2mm]
& &  \hat{F}_3= F_3- B_2\wedge F_1= d\left( C_2 - C_0 B_2\right)= \left[ \partial_\sigma (f_5- f_7 f_4) +\partial_\eta(f_5-f_7 f_4)\right]\wedge \text{Vol}(S^2). \nonumber
\end{eqnarray}
We have defined the Page fluxes $\hat{F}= F \wedge e^{-B_2}$. The associated charges and the flux of $H_3$ must be quantised. Using units such that $\alpha'=g_s=1$ we have,
\begin{equation}
 Q_{Dp,Page}=\frac{1}{(2\pi)^{7-p}}\int_{\Sigma_{8-p}}\hat{F}_{8-p}.\nonumber
\end{equation}
This implies,
\begin{eqnarray}
& & Q_{NS5}=\frac{1}{4\pi^2}\int_{M_3} H_3,\;\;\;\; Q_{D7}= \int_{\Sigma_1} \hat{F}_1,\;\;\; Q_{D5}=\frac{1}{4\pi^2 } \int_{\Sigma_3} \hat{F}_3.\label{pagecharges}
\end{eqnarray}
The cycles $M_3, \Sigma_1, \Sigma_3$ are defined as,
\begin{eqnarray}
& & M_3=[\eta, S^2], \textrm{with}~\sigma=\pm\infty,\;\;\;\;\; \Sigma_1=[\eta], \textrm{with}~\sigma=0, 
\;\;\; \Sigma_3=[\sigma, S^2], \textrm{with}~\eta=\textrm{fixed}.\nonumber
\end{eqnarray}
We allow for a large gauge transformation of $B_2\to B_2 + \Delta d\Omega_2$. This does not change the charge of NS or the D7 brane charge, but as we discuss below has an interesting effect on the 
charge of D5 branes. Let us study the three possible quantised charges.
\\

\noindent\underline{\bf NS-five branes}
\\
Calculating explicitly for the NS five branes charge,
\begin{equation}
\pi Q_{NS5}=\frac{1}{4\pi}\int_{M_3} H_3= \int d\eta \partial_{\eta} f_4(\sigma=\pm\infty,\eta)= f_4(\pm\infty,P)- f_4(\pm\infty,0) .
\end{equation} 
Using now the expressions developed in Appendix \ref{usefulidentities}, in particular eqs.(\ref{B2identity})-(\ref{identityB2}),
we find that the number of NS-five branes satisfies
\begin{equation}
Q_{NS5}= \frac{4}{9\pi} P.\label{QNS5}
\end{equation}
Note that we have included both the contribution of the NS-five branes coming from $\sigma=+\infty$ and those coming from $\sigma=-\infty$. This result suggest that we have $P$ NS-five branes. 
 Now we study the D7 brane charge.
\\

\underline{\bf D7 branes}
\\
For the charge of D7 branes we find,
\begin{eqnarray}
Q_{D7}= \int_{\Sigma_1} \hat{F}_1= \int_{0}^P d\eta \partial_\eta f_7(0,\eta)= f_7(0,P)- f_7(0,0).
\end{eqnarray}
We use the identities developed in Appendix \ref{usefulidentities}, in particular  the identity in eq.(\ref{identityC0}). We have
\begin{equation}
Q_{D7}= 9\left( {\cal R}'(0) -{\cal R}'(P)\right).
\end{equation}
This implies that the function ${\cal R}$ must start and finish as a linear function. That is, in the first interval $0\leq\eta\leq 1$ we must have ${\cal R}= N_1\eta$ and in the last interval
$P-1\leq \eta\leq P$ we must have ${\cal R}= N_P (P-\eta)$, both with integer slopes. As a consequence of this, the number of D7 branes is
\begin{equation}
Q_{D7}= 9 (N_1+ N_{P-1}).\label{numberofd7}
\end{equation} \\

\underline{\bf D5 branes}
\\
For the charge of D5 branes we find, after the large gauge transformation of $B_2\to B_2 + \Delta \text{Vol}(S^2)$,
\begin{eqnarray}
& & \pi Q_{D5}=\frac{1}{4\pi}\int_{\Sigma_3}F_3- (B_2 +\Delta d\Omega_2)\wedge F_1= \int_{-\infty}^{\infty} d\sigma \partial_\sigma \left[f_5- f_7(f_4+\Delta) \right]=\nonumber\\
& & \int_{-\infty}^{-\epsilon} \partial_\sigma [f_5- f_7(f_4+\Delta) ]+\int_{\epsilon}^\infty \partial_\sigma [f_5- f_7(f_4+\Delta) ]
\end{eqnarray}
We need to evaluate the quantity $f_5- f_7(f_4 +\Delta)$ at $\sigma\to\pm\infty$ and $\sigma=\pm \epsilon$ and finally take $\epsilon\to 0$. 
These calculations are streamlined in Appendix \ref{usefulidentities}, see eqs.(\ref{c2b2c01})-(\ref{c2b2c02}). We find that the combination $f_5- f_7 (f_4 +\Delta)$ vanishes at $\sigma\to\pm\infty$. 
This leaves us with
\begin{equation}
\pi Q_{D5}= f_5-f_7(f_4+\Delta) \Big]_{\epsilon}^{-\epsilon}
\end{equation}
evaluated at some fixed value of the $\eta$-coordinate.
Using the expressions in eqs.(\ref{rankfunction}), (\ref{identityC0}) and (\ref{c2b2c02}) we find
\begin{equation}
Q_{D5}=\frac{4}{\pi}\left( {\cal R}(\eta) -{\cal R}'(\eta) \left(\eta-\frac{9\Delta}{4}\right)\right).\label{Qd5}
\end{equation}
This suggests that in each interval the function ${\cal R}(\eta)$ must be linear, with integer slope and intercept. In fact, if we consider  the interval $[k, k+1]$ and choose the large gauge transformation to be interval-dependent $9\Delta=4k$, the rank function is
\begin{equation}
{\cal R}(\eta)= N_k + (N_{k+1}- N_k)(\eta-k), \qquad \eta \in [k,k+1] ,
\end{equation}
we find that the combination in eq.(\ref{Qd5}) gives 
\begin{equation}
Q_{D5}= \frac{4}{\pi} N_k.\label{Qd5final}
\end{equation}
We interpret this as indicating that at each interval, labeled by integer values $[k, k+1]$ of the $\eta$-coordinate, we have a gauge group $SU(N_k)$. The role of the large gauge transformation is to count the D5 charge, without taking into account the charge of D5 induced on the D7 branes.

The values of the Page charges calculated in eqs.(\ref{QNS5}),(\ref{Qd5final}) fail to be integers. This can be remedied by a rescaling, detailed below.
\\

\noindent\underline{\bf Rescaling}
\\
We would like to write our solution in such a way that all the charges are properly quantised. This can be achieved with a redefinition of $V$ and by rescaling the coordinates $(\sigma,\eta)$. These redefinitions introduce overall factors in the fields of eq. \eqref{background}. If we define $V = \nu V_{\text{old}}$ and $(\sigma,\eta)= \mu (\sigma,\eta)_{\text{old}}$ we have the background solving all the equations of motion and SUSY preservation is
\begin{eqnarray}
& & d s^2_{st}\! = \!\mu  d s^2_{\text{old}} , \;\;e^{-2 \Phi} = \frac{\nu^2}{\mu^2} e^{-2 \Phi_{\text{old}}}, \;\; B_2 = \mu B_{2 \, \text{old}} \, , \;\; C_0 = \frac{\nu}{\mu} C_{0 \, \text{old}}, \;\;C_2=\nu C_{2 \, \text{old}}.
\label{rescalingxx}\end{eqnarray}
Here, we choose $\nu = \frac{\pi}{4}$ and $\mu = \frac{9 \pi}{4}$, the fully rescaled configuration is written in eq.(\ref{backgroundrescaled}). In this background we get
\begin{eqnarray}
& & Q_{NS5} = P \, \label{chargesfinal}\\
& &  Q_{D7}[k, k+1]= {\cal R}''(k)=(2 N_{k} - N_{k+1}- N_{k-1}),\; Q_{D7,total}=(N_1+ N_{P-1})= \int_0^P {\cal R}'' (\eta) d \eta ,\nonumber\\
 & & Q_{D5}[k,k+1] = {\cal R}(\eta) -{\cal R}'(\eta) (\eta- \Delta) =N_k\, ,\;\;\;\;\; Q_{D5,total}=\int_0^P {\cal R} ~d\eta.\nonumber 
\end{eqnarray}
In summary, the quantisation of charges, forces us to choose the Rank function in the boundary condition of the PDE, ${\cal R}(\eta)$ of the form,
 \[ {\cal R}(\eta) = \begin{cases} 
          N_1 \eta & 0\leq \eta \leq 1 \\
          N_l+ (N_{l+1} - N_l)(\eta-l) & l \leq \eta\leq l+1,\;\;\; l:=1,...., P-2\\
  %
          N_{P-1}(P-\eta) & (P-1)\leq \eta\leq P . 
       \end{cases}
    \]
The total number of branes in the system is given by eq.(\ref{chargesfinal}). Let us now discuss the associated quantum field theories.

The presence of branes in the configuration, in this case D7 sources, indicates that for our background to be trustable, we need to consider large values of $P$, being the D7 flavour branes separated enough.

\section{Quantum Field Theory}\label{sec:QFT}
Let us briefly discuss some general aspects of five-dimensional QFTs. These theories have symmetries  that constrain enough  the dynamics, allowing analytical treatment of various phenomena. Some of the 5d ${\cal N}=1$ gauge field theories admit a UV fixed point. Conversely, the fix point is deformed by a relevant operator $O\sim \int d^5 x \frac{1}{g^2} F_{\mu\nu}^2$ and in the IR we find a weakly coupled gauge theory description \cite{Seiberg:1996bd}. These fixed points are isolated and strongly coupled \cite{Cordova:2016xhm}, \cite{Chang:2018xmx}. The algebra of minimal 5d SUSY field theories contains eight supercharges and can be derived from the algebra of 6d ${\cal N}=(1,0)$ theories; the five-dimensional SUSY algebra is also related to the algebra of ${\cal N}=2$ theories in four dimensions. This requires an $SU(2)_R$ global R-symmetry. The representations of the algebra include  a hyper-multiplet, with four real scalars and a spinor; a vector multiplet consisting of a vector field, a real scalar and a spinor and a tensor multiplet containing a two form, a real scalar and a fermion (this multiplet is dual to a vector multiplet).

The theories have Coulomb branches, parametrised by the real scalar in the vector multiplet and  Higgs branches described by VEVs for the scalars in the hyper-multiplet. 
There exists a beautiful construction of these branches, based on compactifications of M-theory on three-folds, see for example \cite{Intriligator:1997pq}. The geometric description is very general as it captures SCFTs that do not have  a weakly-coupled description. Also, the description using compactifications of M-theory is powerful enough to determine the  UV flavour symmetry, as this  can enhance
compared to the effective description as a gauge theory. See for example \cite{Apruzzi:2019opn}.

On the other hand, the Lagrangian-based description is useful
as a description of the IR dynamics because some symmetry protected quantities can be calculated and compared against holographic calculations at the UV fix point. Examples of  exact calculations include
the dimension of 'string-like' operators \cite{Bergman:2018hin} in short representations of the global symmetry group (hence protected under RG flow). Other examples include the value of the Partition function for the theory on a five-sphere and Wilson loops. Thanks to the power of localisation--for a review relevant to the present QFTs see \cite{Minahan:2016xwk}--the calculation of these SUSY observables is reduced to finite dimensional integrals, see the papers \cite{Fluder:2018chf}, \cite{Chang:2017mxc}, \cite{Santilli:2021oag}, \cite{Uhlemann:2020bek}, \cite{Uhlemann:2019ypp}. Below, we perform holographic computations that can be put in correspondence with the results of these works. Other calculations, like the dimension of spin-2 operators in the CFT constitute genuine predictions of the holographic set-up \cite{Gutperle:2018wuk}.

To begin with, let us state clearly the correspondence between our Type IIB solutions in Section \ref{sectiongeometry} and the quiver low energy description of the UV SCFTs.

\subsection{Correspondence between our backgrounds and SCFTs}
Now, we put in correspondence our Type IIB backgrounds for particular solutions with five dimensional quiver field theories, that  at high energies flow to super-conformal points (holographically described by the backgrounds). Our proposal is that for a given rank function ${\cal R}(\eta)$
 \[ {\cal R}(\eta) = \begin{cases} 
          N_1 \eta & 0\leq \eta \leq 1 \\
          N_l+ (N_{l+1} - N_l)(\eta-l) & l \leq \eta\leq l+1,\;\;\; l:=1,...., P-2\\
  %
          N_{P-1}(P-\eta) & (P-1)\leq \eta\leq P .
       \end{cases}
    \]
The background constructed following equations (\ref{background}),(\ref{eqfinal})-(\ref{rankfunction}) is the holographic dual of the strong coupling CFT to which a linear quiver theory flows. The gauge group  of the linear quiver is $\Pi_{i=1}^{P-1} SU(N_i)$, the ranks given by the numbers $N_k$ (the rank function evaluated at integer values of the $\eta$-coordinate) and each gauge nodes is connected by bifundamental hypermultiplets. There are also flavour groups $SU(c_i)$ for certain  colour groups. This is indicated by the second derivative of the rank function. The rank of the $k^{th}$ flavour group is 
\begin{equation}
c_k= (2N_k- N_{k+1}- N_{k-1}),
\label{eq:flavor_def}
\end{equation}
as indicated by eq.(\ref{chargesfinal}).

Below, we develop an expression for the Holographic Central Charge, a quantity that measures a weighted version of the internal volume of the manifold in eq.(\ref{background}).
After that, we discuss some 'example case studies'. These examples encapsulate many of the subtleties we wish to discuss in this work. We display the rank function, the associated quiver, the Fourier coefficients and the Potential function $\hat{V}$. We compute the Holographic Central Charge for our case studies finding that the result is proportional to the Free Energy of the associated SCFTs, in the limit  $P\to\infty$, in which our backgrounds are trustable.

\subsection{Holographic Central Charge}
We will calculate the holographic central charge. For a definition appropriate to the type of backgrounds we deal with, we refer the reader to \cite{Macpherson:2014eza},  \cite{Bea:2015fja}.
Briefly summarised, for a holographic background with dilaton, dual to a $d+1$ QFT,
\begin{eqnarray}
ds^2= \alpha(r,\vec{\theta})\left(dx_{1,d}^2 + \beta(r) dr^2 \right) + g_{ij}(r,\vec{\theta}) d \theta^i d \theta^j,\;\;\;\;\; ~~~~e^{-4\Phi}, \label{assignation}
\end{eqnarray}
the weighted internal volume and subsequent definition of the holographic central charge are,
\begin{eqnarray}
& & V_{int}=\int d \vec{\theta}\sqrt{\det[g_{ij}] e^{-4\Phi} \alpha^d},\;\;\;\; H= V_{int}^2,\nonumber\\
& & c_{hol}= \frac{d^d}{G_N}\beta^{d/2}\frac{H^{\frac{2 d+1}{2}}}{(H')^d}\label{chol}
\end{eqnarray}
We apply this formula to the background of eq.(\ref{background})
\begin{eqnarray}
ds^2= f_1 \left[ds^2(\text{AdS}_6) + f_2 d s^2 (S^2) + f_3 (d \sigma^2+d \eta^2)  \right], \;\;\;\;\; e^{-4\Phi}=f_6^2.
\end{eqnarray}
We set  with $d=4$ and use  Poincar\'e coordinates for $AdS_6$,
\begin{eqnarray}
& & ds^2(\text{AdS}_6) = r^2 dx_{1,4}^2+ \frac{dr^2}{r^2},\;\;\;\alpha= f_1(\sigma,\eta) r^2,\;\;\;\;\beta=\frac{1}{r^4}, \\
& & V_{int }= {\cal N} r^4,~~~~~~ {\cal N}=  \int d\theta~d\varphi~ d\sigma~ d \eta~~ \sin\theta~ f_1^4 f_2 f_3 f_6,\nonumber
\end{eqnarray}
which leads to the  expression for the holographic central charge,
\begin{equation}
c_{hol}= \frac{1}{16 G_N}{\cal N}, \label{chol2} 
\end{equation}
where $G_N= 8\pi^6$.
Computing  ${\cal N}$ explicitly we find,
\begin{equation}
{\cal N}= \frac{2^8 \pi}{3} \int_{0}^{P} d \eta \int_{-\infty}^{\infty} d \sigma~\sigma^3 \partial_\sigma V \partial_\eta^2 V .\label{vavar}
\end{equation}
Expanding the derivatives of $V$ in terms of $\hat{V}$ as in \eqref{identities2} and using the expressions in eq.(\ref{identities1}), we are able to express $\mathcal{N}$ as an integral of a double series. Using the orthogonality of the Fourier basis and integrating in $\eta$ reduces the double series to a single one. Performing the integral over $\sigma$ and inserting the final result in eq.(\ref{chol2}) gives,
\begin{equation}
c_{hol}= \frac{1}{2\pi^4}\sum_{k=1}^\infty k a_k^2 . \label{cholfinal}
\end{equation}
This quantity is proportional the Free Energy of the SCFT (calculated in the holographic limit of large $P$), evaluated using matrix models methods in
\cite{Uhlemann:2019ypp}. In particular, we find that the proportionality factor relating ${\cal F}$ in \cite{Uhlemann:2019ypp}
and $c_{hol}$ in eq.(\ref{cholfinal}) is,
\begin{equation}
{\cal F}=-\frac{\pi^6}{4}c_{hol}.\label{conversionU}
\end{equation}
In what follows, we present three case study examples. For each of them we write the rank function ${\cal R}(\eta)$, calculate the Fourier coefficients and the Potential $\hat{V}(\sigma,\eta)$ in eq.(\ref{eq:fourier_vhat}). We draw the associated 5d quiver field theory and evaluate the holographic central charge in eq.(\ref{cholfinal}). 
After that, in Section \ref{secciongeneric} and in  Appendix \ref{centralchargeappendix} we discuss a generic expression.

\subsection{Example 1}\label{example1}

Let us consider a gauge theory, which we call $\tilde{T}_{N,P}$. The gauge theory is described (in the IR) by the quiver 
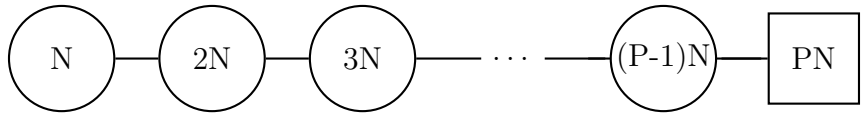
\begin{center}
	\begin{tikzpicture}
	\node (1) at (-6,0) [circle,draw,thick,minimum size=1.4cm] {N};
	\node (2) at (-4,0) [circle,draw,thick,minimum size=1.4cm] {2N};
	\node (3) at (-2,0) [circle,draw,thick,minimum size=1.4cm] {3N};	
	\node (4) at (0,0)  {$\dots$};
	\node (6) at (4,0) [rectangle,draw,thick,minimum size=1.2cm] {PN};
	\node (5) at (2,0)  {(P-1)N};
	\draw[thick] (1) -- (2) -- (3) -- (4) -- (5)-- (6);
	\draw[thick] (2,0) circle (0.7cm) ;
	\draw[thick] (1,0) -- (1.3,0);
	\draw[thick] (2.7,0) -- (3.3,0);
	\end{tikzpicture}\
\end{center}
The rank function associated  to this case is,
\[ {\cal R}(\eta) = \begin{cases} 
N\eta & 0\leq \eta \leq (P-1) \\
N(P-1) (P-\eta)& (P-1)\leq \eta\leq P .
\end{cases}
\]
The number of D7-branes can be read either from $\mathcal{R}'' = NP \delta(\eta-(P-1))$, or from \eqref{chargesfinal} which gives $Q_{D7} = P N$.  The number of D5 branes  at the positions
$\eta=1,2,3,\text{etc}$,  is the value of $\mathcal{R}(\eta)$ at those points.
This coincides with the ranks of the
first, second, third node, etc. In total, we have $\int_0^P {\cal R} d\eta=\frac{NP(P-1)}{2}$ D5 branes.  We also have a total of $P$ NS-five branes.
Given the rank function above, the  coefficient $a_k$ as defined in eq.(\ref{eq:fourier_vhat}) is,
\begin{equation}
a_k= (-1)^{k+1}\frac{N P^3}{k^3 \pi^3} \sin\left( \frac{k\pi }{P}\right). \label{ak1}
\end{equation}
Using eq.(\ref{eq:fourier_vhat}), this leads to the potential $\hat{V}=\sigma V(\sigma,\eta)$
\begin{equation}
\hat{V} = \frac{N P^3}{2 \pi ^3} \text{Re} \left(\text{Li}_3(-e^{-\frac{\pi}{P}  (| \sigma |+i+i \eta  )})-\text{Li}_3(-e^{-\frac{\pi}{P}  (| \sigma |-i+i \eta )}) \right) \, , \label{eq:V_TN}
\end{equation}
which in the large $P$ limit and for $N=1$ matches the one in \ref{subsec:Tn_Uhul}, which is obtained translating \cite{Uhlemann:2020bek} in our formalism.

The holographic central charge for this theory is obtained using eq.(\ref{cholfinal})
\begin{equation}
c_{hol}= \frac{N^2 P^6}{8\pi^{10}}\left(2\zeta(5) -{\text{Li}_5(e^{\frac{2\pi i}{P}} ) -\text{Li}_5(e^{-\frac{2\pi i}{P}})} \right).\nonumber
\end{equation}
This result should only  be trusted for long quivers. Hence, we analyse this result for $P\to\infty$. To leading order we find,
\begin{equation}
c_{hol}=\frac{N^2 P^4}{2\pi^8}\zeta(3) +O\left(\frac{\log P}{P^2}\right).\label{centralexample1}
\end{equation}
Using eq.(\ref{conversionU}), we compare this result  with that of the Free Energy calculated in Section B of \cite{Uhlemann:2019ypp}. Importantly, the scaling with the number of D5 and NS five branes is the same.

\subsection{Example 2}\label{example2}
We consider a second  example, known as the $+_{P,N}$ theory. The rank function is defined as,
\[ {\cal R}(\eta) = \begin{cases} 
N\eta & 0\leq \eta \leq 1 \\
N & 1\leq \eta\leq (P-1)\\
N (P-\eta) & (P-1)\leq \eta\leq P .
\end{cases}
\]
We have $N$ D7-branes  localised at $\eta = 1$ and $N$ D7 branes at $\eta= P-1$. This follows from $\mathcal{R}'' = N \delta(\eta-1)+N \delta(\eta-(P-1))$. There are a total of $(P-1)N$ D5-branes, as calculated by $\int_0^P {\cal R} d\eta$. The number comes from $N$ D5 branes for each integer value of $\eta$ between $[1,P-1]$. This is equivalent to a linear quiver field theory,
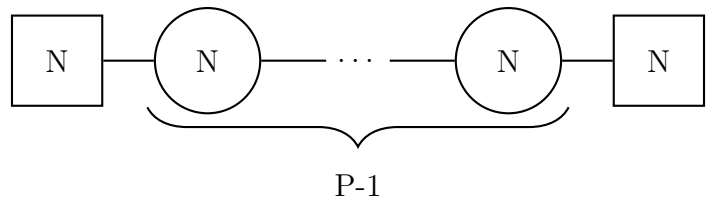
\begin{center}
	\begin{tikzpicture}
	\node (1) at (-4,0) [rectangle,draw,thick,minimum size=1.2cm] {N};
	\node (2) at (-2,0) [circle,draw,thick,minimum size=1.4cm] {N};
	\node (3) at (0,0)  {$\dots$};
	\node (5) at (4,0) [rectangle,draw,thick,minimum size=1.2cm] {N};
	\node (4) at (2,0) [circle,draw,thick,minimum size=1.4cm] {N};
	\draw[thick] (1) -- (2) -- (3) -- (4) -- (5);
	\draw [decorate,decoration={brace,amplitude=15pt,mirror},thick,yshift=-1.5em]
	(-2.8,0) -- (2.8,0) node[midway,yshift=-2.5em]{P-1};
	\end{tikzpicture}
\end{center}
Using eq.(\ref{eq:fourier_vhat}) we calculate
\begin{equation}
a_k= \frac{N P^2}{k^3 \pi^3} \sin\left( \frac{k\pi }{P}\right) \left(1+(-1)^{k+1}\right),\label{ak2}
\end{equation}
which leads to the potential
\begin{align}
\hat{V}=& \frac{N P^2}{2 \pi ^3} \text{Re}\big(\text{Li}_3(e^{-\frac{\pi}{P}  (| \sigma |-i\eta +i)}) -\text{Li}_3(-e^{-\frac{\pi}{P}  (| \sigma |-i\eta +i)}) \nonumber\\[2mm]
+&\text{Li}_3(-e^{-\frac{\pi}{P}  (| \sigma |-i\eta -i)})-\text{Li}_3(e^{-\frac{\pi}{P}  (| \sigma |-i\eta -i)}) \big) \, . \label{eq:+MN}
\end{align}
Notice that this potential in the large $P$ limit, matches the one in Appendix  \ref{subsec:+MN_Uh}.
The holographic central charge, using eq.(\ref{cholfinal}) and the Fourier coefficient in eq.(\ref{ak2}) is
\begin{equation}
c_{hol}= \frac{N^2 P^4}{32\pi^{10}}\left(31\zeta(5) + 8{\text{Li}_5(-e^{\frac{2\pi i}{P}} ) +8 \text{Li}_5(-e^{-\frac{2\pi i}{P}})}  -   8{\text{Li}_5(e^{\frac{2\pi i}{P}} ) -8\text{Li}_5(e^{-\frac{2\pi i}{P}})} \right).\nonumber
\end{equation}
Using that this result is only  trustable for long quivers, we analyse it for $P\to\infty$. To leading order we find,
\begin{equation}
c_{hol}=\frac{7 N^2 P^2}{4\pi^8}\zeta(3) +O\left(\frac{\log P}{P^2}\right).\label{centralexample2}
\end{equation}
Using eq.(\ref{conversionU}), we compare this result  with that of the Free Energy calculated in Section A of \cite{Uhlemann:2019ypp}. Importantly the scaling with the number of D5 and NS five branes is the same.

\subsection{Example 3}\label{example3}
This example is inspired on the material in Section F of \cite{Uhlemann:2019ypp}.
The rank function is,
\[ {\cal R}(\eta) = \begin{cases} 
\left[N- j K + (j-1) \right] \eta & 0\leq \eta \leq 1 \\
(N-j K)+ (j-1) \eta & 1\leq \eta\leq K\\ 
N-\eta& K\leq \eta\leq N 
\end{cases}
\]
We have two stacks of D7 branes. The first stack is located at $\eta=1$ and contains $(N-jK)$ D7 branes. The second stack, located at $\eta=K$ has $j$ D7 branes. This is read from  
\begin{equation}
R''= (N- j K)\delta(\eta-1)+ j \delta(\eta - K).\nonumber
\end{equation} 
There are $P$ NS-five branes and  $\frac{1}{2} \left[ N(N-1) + jK(K-1)\right]$ D5 branes in the background. 
\\
The quiver associated with this rank function is,
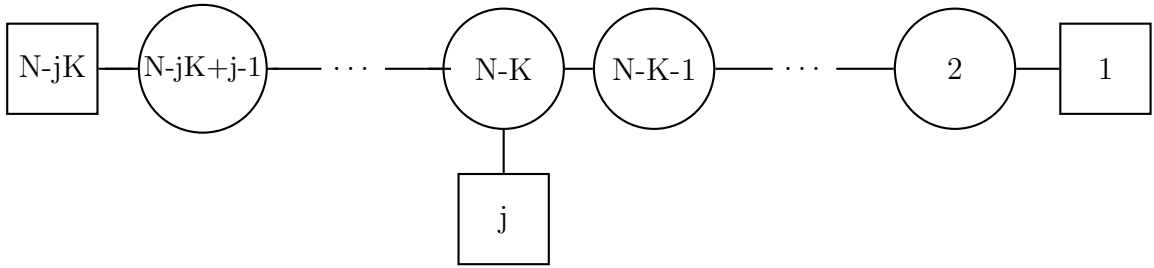
\begin{center}
	\begin{tikzpicture}
	\node (1) at (-4,0) [rectangle,draw,thick,minimum size=1.2cm] {N-jK};
	\node (2) at (-2,0)  {\small N-jK+j-1};
	\node (3) at (0,0)  {$\dots$};
	\node (4) at (2,0) [circle,draw,thick,minimum size=1.6cm] {N-K};
	\node (5) at (4,0) [circle,draw,thick,minimum size=1.6cm] {N-K-1};
	\node (6) at (6,0)  {$\dots$};
	\node (7) at (8,0) [circle,draw,thick,minimum size=1.6cm] {2};
	\node (8) at (10,0) [rectangle,draw,thick,minimum size=1.2cm] {1};
	\node (9) at (2,-2) [rectangle,draw,thick,minimum size=1.2cm] {j};
	\draw[thick] (1) -- (2) -- (3) -- (4) -- (5) --(6)--(7)--(8);
	\draw[thick] (4) -- (9);
	(-2.8,0) -- (2.8,0) node[midway,yshift=-2.5em]{P-2};
	\draw[thick] (-2,0) circle (0.85cm) ;
	\draw[thick] (1,0) -- (1.3,0);
	\draw[thick] (-2.85,0) -- (-3.3,0);
	\draw[thick] (-1.15,0) -- (-1,0);
	\end{tikzpicture}
\end{center}
Note that the rank of the $ l^{th}$-gauge group is precisely the value of the rank function at $\eta=l$. 
The Fourier coefficient $a_m$ computed with eq.(\ref{eq:fourier_vhat}) is
\begin{equation}
a_m=\frac{N^2}{m^3 \pi^3} \left[ (N-j K) \sin \left(\frac{\pi  m}{N}\right)+j \sin \left(\frac{\pi  K m}{N}\right)   \right].\label{ak3}
\end{equation}
Using eq.(\ref{eq:fourier_vhat}), this leads to the potential
\begin{align}
\hat{V}=& \frac{N^2}{4 \pi ^3} \text{Re}\Big[j \big(\text{Li}_3(e^{-\frac{\pi}{N}  ( |\sigma| - i K+i \eta  )})-\text{Li}_3(e^{-\frac{\pi}{N}  ( |\sigma| + i K+i \eta  )}) \big) \nonumber \\[2mm]
+& (N-j K) \big(\text{Li}_3(e^{-\frac{\pi}{N}  ( |\sigma| - i+i \eta  )})-\text{Li}_3(e^{-\frac{\pi}{N}  ( |\sigma| + i+i \eta  )})\big)\Big] .\label{eqVcompl}
\end{align}
Below we will compare the Holographic Central Charge computed using the background with a field theoretical result for this SCFT.

The present example is more demanding than the previous ones. 
Applying equation (\ref{cholfinal}) with the Fourier coefficients in eq.(\ref{ak3}) gives a very involved result. Consider the case for which $K=l N$ and $N\to\infty$ keeping $j$ finite. To leading order in $N$ we find,
\begin{eqnarray}
& & c_{hol}=
\frac{N^4}{2\pi^8} \Big[ \frac{j^2}{2\pi^2} \left(\zeta(5) - \frac{\text{Li}_5(e^{2 i\pi l}) +\text{Li}_5(e^{-2 i\pi l})}{2}\right) 
+ (j l-1)^2 \zeta(3) +\nonumber\\
& & 
\frac{2j (1-j l)}{\pi} \left(\frac{\text{Li}_4(e^{i l \pi} ) - \text{Li}_4(e^{- i l \pi})}{2i}\right)\Big].\label{centralexample3}
\end{eqnarray}
Using eq.(\ref{conversionU}), this result should be compared with  equation (4.61) of \cite{Uhlemann:2019ypp}.
\subsection{General case}\label{secciongeneric}
After these examples, we discuss the general case. Since the procedure is identical to the one in the previous section, we relegate all the details about this section to Appendix \ref{centralchargeappendix}.
The most general rank function is given by
\[ {\cal R}(\eta) = \begin{cases} 
N_1 \eta & 0\leq \eta \leq 1 \\
N_l+ (N_{l+1} - N_l)(\eta-l) & l \leq \eta\leq l+1,\;\;\; l:=1,...., P-2\\
%
N_{P-1}(P-\eta) & (P-1)\leq \eta\leq P .
\end{cases}
\]
which correspond to the following quiver theory
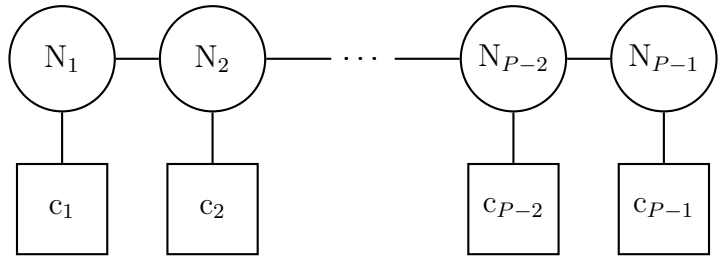
\begin{center}
	\begin{tikzpicture}
	\node (1) at (-4,0) [circle,draw,thick,minimum size=1.4cm] {N$_1$};
	\node (2) at (-2,0) [circle,draw,thick,minimum size=1.4cm] {N$_2$};
	\node (3) at (0,0)  {$\dots$};
	\node (5) at (4,0) [circle,draw,thick,minimum size=1.4cm] {N$_{P-1}$};
	\node (4) at (2,0) [circle,draw,thick,minimum size=1.4cm] {N$_{P-2}$};
	\draw[thick] (1) -- (2) -- (3) -- (4) -- (5);
	\node (1b) at (-4,-2) [rectangle,draw,thick,minimum size=1.2cm] {c$_1$};
	\node (2b) at (-2,-2) [rectangle,draw,thick,minimum size=1.2cm] {c$_2$};
	\node (3b) at (0,0)  {$\dots$};
	\node (5b) at (4,-2) [rectangle,draw,thick,minimum size=1.2cm] {c$_{P-1}$};
	\node (4b) at (2,-2) [rectangle,draw,thick,minimum size=1.2cm] {c$_{P-2}$};
	\draw[thick] (1) -- (1b);
	\draw[thick] (2) -- (2b);
	\draw[thick] (4) -- (4b);
	\draw[thick] (5) -- (5b);
	\end{tikzpicture}
\end{center}
where $c_i$ are defined in \eqref{eq:flavor_def}. By convention $N_0=N_P=0$. The Fourier coefficients $a_k$ are given by 
\begin{equation}
a_k=\frac{P^2}{\pi^3 k^3}\sum_{s=1}^{P-1} c_s \sin \left( \frac{k\pi s}{P}\right)
\end{equation}
and define the following potential
\begin{equation}
\hat{V}= \frac{P^2}{2\pi^3}\sum_{s=1}^{P-1} c_s \text{Re} \left(\text{Li}_3(e^{-\frac{\pi (|\sigma |+ i\eta- i s)}{P}})-\text{Li}_3(e^{-\frac{\pi (|\sigma |+ i\eta+ i s)}{P} }) \right) .
\end{equation}

Interestingly, the holographic central charge can still be computed analytically in the generic case, and leads to the following result:
\begin{eqnarray}
c_{hol}=  -\frac{P^4}{4\pi^{10}} \sum_{l=1}^{P-1} \sum_{s=1}^{P-1} c_l c_s \text{Re}\left(  \text{Li}_5( e^{i\frac{\pi (l+s)}{P}}) - \text{Li}_5( e^{i\frac{\pi (l-s)}{P}}) \right).
\end{eqnarray}
The details about this computation and the comparison of the general result with the previous three example, can all be found in appendix \ref{centralchargeappendix}. The trustability of these generic backgrounds is also subject to the $P\to\infty$ and well-separated flavour groups condition. Since the rank function starts and end in zero, this result matches exactly the general SCFT free energy (3.17) in \cite{Uhlemann:2019ypp}.

We now discuss another observable that can be used as a check of the correspondence proposed here.

\subsection{Wilson Loops}
In this section we discuss Wilson loops, focusing on those in antisymmetric representations. We follow the lead of \cite{Uhlemann:2020bek}. As discussed there the relevant object to compute these Wilson loops is a probe D3 brane extended over AdS$_2$ (inside AdS$_6$) and the two sphere. We also need to switch  electric and magnetic fluxes $\hat{f}_2$ on the brane. Writing
\begin{equation}
ds^2(\text{AdS}_6)= - \cosh^2\rho \, ds^2 (\text{AdS}_2) + d\rho^2+ \sinh^2\rho \, ds^2 (S^3),\;\;\; \hat{f}_2={\cal F}_e\text{Vol}(\text{AdS}_2) + {\cal F}_m \text{Vol}(S^2) \nonumber
\end{equation}
where the D3-brane sits at $\rho=0$.
The metric-flux induced  on the brane and the D3-brane action $S_{D3}= S_{BI}+ S_{WZ}$ are
\begin{eqnarray}
& & ds_{ind}^2= f_1 ds^2 (\text{AdS}_2)+ f_1 f_2 ds^2(S^2), \quad  {\cal F}= B_2+\hat{f}_2= {\cal F}_e \text{Vol}(\text{AdS}_2)+(f_4 +{\cal F}_m ) \text{Vol}(S^2) \, , \nonumber\\[2mm]
& &S_{BI}=\int   \sqrt{e^{-2\Phi}\det[g+ {\cal F}] }= 4 \pi T_{D3}\text{vol}(\text{AdS}_2)\sqrt{ f_6 (f_1^2+{\cal F}_e^2)\left( f_1^2f_2^2 +(f_4 +{\cal F}_m)^2\right)},\nonumber\\
& & S_{WZ}= - T_{D3}\int C_p\wedge  e^{-{\cal F}}= -4 \pi T_{D3}\text{vol}(\text{AdS}_2) \left( f_5- f_7(f_4 +{\cal F}_m)\right), \label{actionD3} 
\end{eqnarray}
where $\text{vol}(\text{AdS}_2)=-2 \pi$ is the (renormalized) AdS$_2$ volume.
We follow \cite{Uhlemann:2020bek} for the particular values of ${\cal F}_e, {\cal F}_m$ needed to implement SUSY on the D3 probe while the position of the D3 in $(\eta,\sigma)$ is not fixed by any condition, which means that we have a two-parameters family of solutions. Calculating the Legendre transform of the D3 action, we find for the VEV of the Wilson loop
\begin{equation}
\log \langle W \rangle= S_{D3}-\frac{\delta S_{D3}}{\delta {\cal F}_e} \sim \sigma^2\partial_\sigma V= (\sigma\partial_\sigma \hat{V}-\hat{V})\label{wilsonfinal}
\end{equation}
In what follows, we work up to a proportionality factor. Using the expressions in the appendices, in particular eqs.(\ref{identities1}),(\ref{akgen2}),(\ref{c_s_def}), we have
\begin{eqnarray}
& & -\log \langle W \rangle \sim \hat{V}-\sigma \partial_\sigma\hat{V}= \sum_{k=1}^{\infty} a_k \sin\left(\frac{k\pi \eta}{P}\right) e^{-\frac{k\pi |\sigma|}{P}} \left(1+\frac{k\pi |\sigma|}{P}\right)\label{wilsonfinalex}\\
& & =\frac{P^2}{4\pi^3}
\sum_{s=1}^{P-1} c_s \sum_{k=1}^{\infty} \left(\frac{1}{k^3} +\frac{\pi |\sigma|}{P k^2} \right) \text{Re}\left[ e^{ -\frac{k\pi}{P} (|\sigma|+ i(\eta+s)) }  - e^{ -\frac{k\pi }{P}(|\sigma|+ i(\eta-s))}\right]\nonumber
\end{eqnarray}
which is
\begin{eqnarray}
& &\log \langle W \rangle \sim  -\frac{P^2}{4\pi^3}
\sum_{s=1}^{P-1} c_s \text{Re} \Big[\text{Li}_3(e^{-\frac{\pi}{P} (|\sigma|-i(\eta+s))}) - \text{Li}_3(e^{-\frac{\pi}{P} (|\sigma|-i (\eta-s))}) \nonumber \\
& & \qquad \qquad \, \, \, +\frac{\pi |\sigma|}{P}\big(\text{Li}_2(e^{-\frac{\pi}{P} (|\sigma|-i(\eta+s))}) - \text{Li}_2(e^{-\frac{\pi}{P} (|\sigma|-i (\eta-s))}) \big) \Big].\label{wilsonloopfinal}
\end{eqnarray}
As in the previous sections, this expression should be compared with results on the SCFT side, derived with matrix model methods. Such comparison is meaningful only for large values of $P$.

Notice that, for the small $s$ and $P-s$, the large $P$ limit changes the contribution on the summation
\begin{eqnarray}
& \frac{s}{P} \to 0 & \quad -\frac{P^2}{4\pi^3} c_s s P \,  \text{Im} \left[ \frac{\pi |\sigma|}{P} \log (1-e^{-\frac{\pi}{P}  (| \sigma | +i \eta )}) -\text{Li}_2 (e^{-\frac{\pi}{P}  (| \sigma | +i \eta )})\right] \, , \nonumber\\
& \frac{s}{P} \to 1 & \quad \frac{P^2}{4\pi^3} c_s (P-s) P \,  \text{Im} \left[ \frac{\pi |\sigma|}{P} \log (1+e^{-\frac{\pi}{P}  (| \sigma | +i \eta )}) -\text{Li}_2 (-e^{-\frac{\pi}{P}  (| \sigma | +i \eta )})\right] \, , \nonumber
\end{eqnarray}
While if $s/P$ is finite, no simplification occurs. Let us see these formulas working explicitly in our examples of Sections \ref{example1}, \ref{example2}, \ref{example3}.
\\

\noindent\underline{\bf Example 1}
\\
In this case $c_s= N P \delta_{s,P-1}$. It is convenient to introduce the notation
\begin{equation}
z= -\xi e^{\frac{i \pi}{P}},\;\; w= -\xi e^{-\frac{i\pi}{P}}, \;\;\;\;\xi= e^{-\pi\frac{(|\sigma| +i\eta)}{P}},\;\;\;\; |\sigma|= -\frac{P}{\pi}\log|\xi|.\label{zwxi}
\end{equation}
Evaluating explicitly eq.(\ref{wilsonloopfinal}), we find
\begin{equation}
-\log \langle W \rangle \sim-\frac{NP^3}{4\pi^3}\text{Re}\left[  \text{Li}_3(z) -\text{Li}_3(w) + \log|\xi| ( \text{Li}_2(z) -\text{Li}_2(w)) \right].
\end{equation}
Now we expand the expression above for $P\to\infty$. In this expansion it is important to keep in mind that $\frac{|\sigma|}{P},\frac{\eta}{P}$ are taken to be fixed, hence the expansion acts on the $e^{\frac{i\pi}{P}}$ factors. This leads to,
\begin{eqnarray}
\log \langle W \rangle \sim -\frac{NP^2}{\pi^2}\left[\text{Im} ( \text{Li}_2(-\xi) ) + \log |\xi|\text{Arg}(1+ \xi) \right].\label{wilsonexample1}
\end{eqnarray}
This result is to be compared (for $N=1$)  with that in eq.(4.36) of the paper \cite{Uhlemann:2020bek}.
\\

\noindent\underline{\bf Example 2}
\\
In this case we have $c_s=N(\delta_{s,1} + \delta_{s,P-1})$. The calculation proceeds similarly as that described above, in particular, similar definitions as those of eq.(\ref{zwxi}) are used and eq.(\ref{wilsonloopfinal}) gives,
\begin{eqnarray}
& & -\log \langle W \rangle \sim -\frac{NP^2}{4\pi^3}\text{Re}\Big[  \text{Li}_3(\xi e^{-\frac{i\pi}{P}}) -\text{Li}_3(\xi e^{\frac{i\pi}{P}}) + \text{Li}_3(-\xi e^{\frac{i\pi}{P}}) -  \text{Li}_3(-\xi e^{-\frac{i\pi}{P}}) \Big] \nonumber\\
& &+ \frac{NP^3}{4\pi^3}\log|\xi| \text{Re} \Big[  \text{Li}_2(\xi e^{-\frac{i\pi}{P}}) -\text{Li}_2(\xi e^{\frac{i\pi}{P}}) + \text{Li}_2(-\xi e^{\frac{i\pi}{P}}) -
 \text{Li}_2(-\xi e^{-\frac{i\pi}{P}}) \Big].\nonumber
\end{eqnarray}
As in all previous holographic calculations, they should be trusted only for $P\to\infty$, keeping fixed  $\frac{|\sigma|}{P},\frac{\eta}{P}$. This gives
\begin{equation}
-\log \langle W \rangle \sim \frac{NP}{4\pi^2} \text{Im}\left[ \text{Li}_2(-\xi) -\text{Li}_2(\xi)\right] -\frac{NP}{\pi^2}\log|\xi |\left[ \text{Arg}(1-\xi) -\text{Arg}(1+\xi)\right] . \label{wilsonexample2}
\end{equation}
\\

\noindent\underline{\bf Example 3}
\\
We describe briefly this example as it is a combination of the two examples above, with the interesting subtlety of the scaling $K= l N$, as we did in Section \ref{example3}). After some algebra we find the expression for $P\to\infty$,
\begin{eqnarray}
& & -\log \langle W \rangle \sim  \frac{N^2}{\pi^2}(j l-1) \left[ \text{Im} \text{Li}_2(\xi) - \log|\xi | \text{Arg}(1-\xi)  \right]\label{ex3wilsonfinal}\\
& & -\frac{N^2 j}{2\pi^3} \Big[\log|\xi | \text{Re}( \text{Li}_2(\xi e^{-i l \pi}) -\text{Li}_2(\bar{\xi} e^{-i l\pi}) ) +\text{Re}(\text{Li}_3(\xi e^{-i l \pi}) -\text{Li}_3(\bar{\xi} e^{-i l\pi}))    \Big].\nonumber
\end{eqnarray}
These examples show that our calculations (valid for the SCFT) are capturing the same Physics as the matrix models do, for the IR-QFT description of the system \cite{Uhlemann:2020bek}, \cite{Uhlemann:2019ypp}, \cite{Santilli:2021oag}.

Let us now change direction and discuss some solutions of the PDE (\ref{diffeq}),(\ref{eqfinal}) that do not satisfy the boundary conditions in eq.(\ref{bc}). In some cases of interest, we provide a physical understanding of these backgrounds.

\section{Some special solutions}\label{section-special}
So far we have discussed the solutions obtained in Section \ref{resPDE} where the potential $V$ was expanded in terms of its Fourier modes as in equation (\ref{eq:fourier_vhat}). 
In this section we expand $V$ in Taylor series and we keep only some terms in the expansion (the purpose of this will become clear below). In fact, consider the case $ \sigma\geq 0$ for simplicity, we can write
\begin{eqnarray}
V(\sigma,\eta)= \sum_{k=1}^\infty a_k \frac{e^{-\frac{k\pi \sigma}{P}}}{\sigma} \sin\left( \frac{k \pi\eta}{P}\right)=
 \sum_{k=1}^\infty \sum_{n=0}^\infty \sum_{l=0}^\infty \frac{ (-1)^{l+n}a_k}{l! (2n+1)!}
\left(\frac{k\pi}{P}\right)^{2n+l+1}\sigma^{l-1}\eta^{2n+1}.\label{Vapproxs}
\end{eqnarray}
Each term in the sum, weighted by a power of $\left(\frac{k\pi}{P}\right)$, is a solution to eq.(\ref{diffeq}) even if it does not satisfy the boundary conditions in (\ref{bc}). We refer to them as a  `partial polynomial solution'.

This potential can actually be generalised considering the most general solution to equations (\ref{diffeq}),(\ref{eqfinal}) without imposing the boundary conditions in eq.(\ref{bc}).
In fact, consider partial polynomials of the form
\begin{equation}
\label{eq:V_poly}
V = \sum_{n=1}^\infty \sum_{k=0}^{\lfloor \frac{n-1}{2} \rfloor} (-1)^k \sigma^{n-2k-2} \eta^{2k} \left[a_n \binom{n}{2k+1} \eta + b_n \binom{n-1}{2k}\right].
\end{equation}
These solutions are obtained  by taking the generic potential $\hat{V}(\sigma,\eta)$ that according to eq.(\ref{eqfinal}) is a harmonic function. The most general power series solution can be written as the real or imaginary part of a polynomial in the complex variable $z = \sigma + i \eta$. The expression  in eq.(\ref{eq:V_poly}) is the most general potential $V$, 
where the coefficients $b_n$ are given by the real part of $z^{n-1}$ while the  $a_n$ are the imaginary part of $z^n$. Indeed, notice that the potential in eq.(\ref{Vapproxs}) is a particular case of that in eq.(\ref{eq:V_poly}) with $b_n=0$, as one can check after exchanging $n\to k$ and $2n+l+1\to n$.

These backgrounds do not have a good interpretation in terms of a long linear quiver, however some truncations contained in eq.(\ref{eq:V_poly}) reproduce previously know solutions. 
In particular, we find below the Abelian and non-Abelian T-dual of the unique SUSY AdS$_6$ solution in massive type IIA  \cite{Brandhuber:1999np}. 

\subsection{Type IIA Abelian T-dual}\label{subsectionATD}
The Abelian T-dual of the D4/D8 system in type IIA can be obtained by setting $a_n=0$ in eq.\eqref{eq:V_poly} and truncating the series to $n=4$. The $b_2$ term is a constant and is set to zero. 
We have to set $b_3=0$ for the background obtained in Type IIB to have the U(1) symmetry necessary to T-dualise back to massive IIA. 
Summarising, the solution is defined by the potential
\begin{equation}
\label{eq:abelian_V}
V_{ATD} = \frac{b_1}{\sigma }+b_4 \left(3 \eta ^2-\sigma ^2\right) \, . 
\end{equation}
In order to match the conventions of \cite{Lozano:2018pcp} with $L=1$, we set
\begin{equation}
b_1 = \frac{81}{512} \, , \qquad b_4 = -\frac{m}{486}.
\end{equation}
We also perform the  change of coordinates
\begin{equation}
\sigma = \frac{27}{8} W^2  \cos \alpha \, , \qquad \eta = \frac{9 }{2} \psi \, , \qquad W=(m \cos \alpha)^{-\frac{1}{6}} \, .\label{changeofcoordinates4}
\end{equation}
After these redefinitions we find the fluxes and dilaton,
\begin{equation}
F_1 = -m d \psi \, , \quad H_3 = d \psi \wedge \text{Vol}(S^2) \, , \quad F_3 = - \frac{5}{8 W^2} \sin^3 \psi d \psi \wedge \text{Vol}(S^2),  \, \quad e^{-\Phi} = \frac{3 \sin \alpha}{4 W^4}
\end{equation}
while the metric is
\begin{equation}
 d s^2 = \frac{W^2}{4} \left[9 ds^2(\text{AdS}_6) + \sin^2 \alpha ds^2(S^2)\right]+ \frac{4}{W^2 \sin^2 \alpha} \left(d \psi^2 + \frac{W^4}{4} \sin^2 \alpha d \alpha^2 \right) \, .
\end{equation}

\subsection{Type IIA non-abelian T-dual}\label{subsectionNATD}
Another solution of interest is the non-Abelian T-dual of the type IIA D4/D8 system. This was well studied in \cite{Lozano:2012au}, \cite{Lozano:2013oma}, \cite{Lozano:2018pcp}. This background can be obtained from eq.\eqref{eq:V_poly} truncating  the expansion to $n=4$, with $b_n=0$.  Conversely, by considering the partial polynomials in eq.(\ref{Vapproxs}) and truncating  to $k=4$. We can set $a_2=0$ without loosing generality.
 The solution of  \cite{Lozano:2012au}, \cite{Lozano:2013oma}, \cite{Lozano:2018pcp} is recovered imposing  $a_3=0$. The potential for this case reads
\begin{equation}
V_{NATD} = \frac{a_1 \eta }{\sigma }+ a_4 \left(4\eta  \sigma ^2-4\eta ^3\right) \, .\label{VNATD}
\end{equation}
Comparing with eq.(\ref{eq:abelian_V}), we notice that $\frac{d }{d\eta} V_{NATD}\sim V_{ATD}$. In order to match the conventions of \cite{Lozano:2018pcp}, we set
\begin{equation}
a_1 = \frac{1}{128} \, , \qquad a_4 = \frac{m}{432}.
\end{equation}
We also perform the change of coordinates
\begin{equation}
\eta = r \, , \qquad \sigma = \frac{3}{4} W^2  \cos \alpha \, ,
\end{equation}
where $W$ is defined in eq.(\ref{changeofcoordinates4}).
The fluxes and dilaton are given by
\begin{eqnarray}
& & F_1 = -m r d r - \frac{5}{8 W^2}  \sin ^3 \alpha d \alpha \, , \quad B= \frac{2r^3}{9\Delta} \text{Vol}(S^2) \, , \quad  e^{-\Phi} = \frac{3 \sin \alpha}{4W^4} \sqrt{\Delta}  \nonumber\\
& & F_3 = \frac{2}{9}\frac{r^2 \sin^3 \alpha}{16 \Delta  W^2 \cos \alpha} ( \sin \alpha d r -10 r \cos \alpha d \alpha ) \wedge \text{Vol}(S^2),
\;\;\;\;
\Delta = r ^2+ \left(\frac{ W \sin \alpha}{2}\right)^4 \qquad \label{fluxesdilatonNATD}
\end{eqnarray}
whilst the metric reads\footnote{Notice that in this case, in order to better match  the conventions of \cite{Lozano:2018pcp}, it is necessary to perform a rescaling of the background as in \eqref{eq:rescaling_10D} with $a^2 = \frac92$.}:
\begin{equation}
d s^2 = \frac{W^2}{18} \left[9 ds^2(\text{AdS}_6) + \frac{r^2}{\Delta} \sin^2 \alpha ds^2(S^2)\right]+ \frac{8}{9 W^2 \sin^2 \alpha} \left(d r^2 + \frac{W^4}{4} \sin^2 \alpha d \alpha^2 \right) \, .\label{metricNATD}
\end{equation}
\subsection{ What do we learn: SCFTs resolving gravity singularities}
The SUSY AdS$_6$ background in massive IIA  \cite{Brandhuber:1999np}   is the holographic dual to the five dimensional $USp(2N)$ theory with one anti-symmetric multiplet and  $N_f<8$ hypermultiplets. The T-dual version of that solution, discussed in Section \ref{subsectionATD},  is holographically dual to the same CFT, in virtue of T-duality producing an exactly equivalent background  from the perspective of a string.

The same reasoning cannot be applied to non-Abelian T-duality, that should be thought of as a solution generating technique (in contrast to a symmetry of the string theory). A natural question is then how to interpret holographically the solution generated by non-Abelian T-duality. This problem was tackled in a variety of examples in diverse dimensions. See for example
\cite{Lozano:2016kum},  \cite{Lozano:2016wrs}, \cite{Lozano:2017ole},  \cite{Itsios:2017cew}. The idea in those papers is that given a background dual to a well defined CFT, a particular zoom-in (or a Penrose-like scaling) reveals the presence of a solution obtained via non-Abelian T-duality. In our case we find analogous behaviour. 

In fact, the analysis of the background in eqs.(\ref{fluxesdilatonNATD})-(\ref{metricNATD}) reveals that it is highly singular. The solution is not trustable and a holographic interpretation is difficult to propose. For example, given the potential in eq.(\ref{VNATD}), we can calculate the associated Rank function, that in this case is 
\begin{equation}
R(\eta)= \partial_\sigma \hat{V}|_{\sigma=0}= \partial_\sigma\left(\sigma V\right)|_{\sigma=0}= -\frac{m}{108} \eta^3.
\end{equation}
With such rank function we would have various problems, for example in defining gauge groups. Also,  there is no bound for the $\eta$-coordinate, which would suggest an infinite number of gauge groups.  This together with the various singularities of the background in eqs.(\ref{fluxesdilatonNATD})-(\ref{metricNATD}) make a holographic interpretation quite hard to propose.

Nevertheless, there is a cure for these problems. Indeed, consider a  5d SCFT described by a rank function and potential function ${V}(\sigma,\eta)$ in eqs.(\ref{eq:fourier_vhat}) and (\ref{Vapproxs}). This potential (and a suitable Rank function) describe holographically a well defined SCFT.  Scaling this potential, keeping terms below order $O(\frac{1}{P^5})$, and restricting to the case in which $a_3=0$,  we find the potential function $V(\sigma,\eta)$ in eq.(\ref{VNATD}), characterising the non-Abelian T-dual of the SUSY AdS$_6$ background of  massive IIA. This suggests the idea that the backgrounds obtained using non-Abelian T-duality, should be thought of as 'slices'  (or a zoom-in) of a more general solution. The infinite family of solutions in eqs.(\ref{eq:fourier_vhat}), (\ref{Vapproxs}) have a nice holographic dual, but each 'slice'  does not, hence requiring a completion. This completion is what the full solutions in eqs.(\ref{eq:fourier_vhat}) and (\ref{Vapproxs}) provide for the solution in eq(\ref{VNATD}).
For the case at hand, this was proposed in  \cite{Lozano:2018pcp}. In Section \ref{subsectionNATD} we reproduce this, easily expressing it within our formalism.

\section{Conclusions}\label{concl}
Let us start with a brief summary of the contents of this work.

Building on the work of \cite{AGLMZ}, in Section \ref{sectiongeometry}, we run the algorithm described in the introductory section and wrote a family of Type IIB configurations. This infinite family of backgrounds is fully described in terms of a potential function $V(\sigma,\eta)$ solving a linear PDE with appropriate boundary conditions.
Our choice of boundary conditions is motivated by the quantisation of the  Page charges and allows for a clean identification with a quiver field theory. This quiver is UV completed by a SCFT, dual to our Type IIB background.
We analyse the behaviour of the solutions at special points in the $(\sigma,\eta)$-plane, revealing the presence of 'colour-branes' dissolved in fluxes and 'flavour branes' present in the background as a localised physical object.

In Section \ref{sec:QFT} we write a precise  correspondence between the SCFT, the Rank function used as boundary condition and the string background. Pedagogical examples are discussed giving detailed expressions for the Rank function, the potential function $V(\sigma,\eta)$ and various observables. These protected observables act as tests of the correspondence we proposed. General expressions for the potential function, the central charge and the Wilson loops VEVs, are provided for the reader wishing to attempt their own example as a particular case.

In Section \ref{section-special}, special backgrounds that do not strictly satisfy the healthy boundary conditions are discussed. In particular the singular background obtained by non-Abelian T-duality of the Brandhuber-Oz solution in massive IIA. We gave a field theoretical mechanism to cure the singular behaviour, by embedding it into healthy solutions as those discussed in the other sections.

This paper suggests various lines for future development. First of all, we have so far considered the holographic dual of balanced quivers only. However, it seems possible to generalized our formalism to larger class of SCFT by changing the electrostatic problem considered in figure \ref{fig:elect_problem}. In particular, it looks like there is a close correspondence between the electrostatic problem for $\partial_\eta^2 (\sigma V)$\footnote{Basically, the second derivative respect to $\eta$ of eq.\eqref{eq:electrostatic_problem_2}.} and the one considered in \cite{Uhlemann:2019ypp} on the CFT side.

Moreover, it would be good to match various  other results obtained for different observables as the ones in  \cite{Bergman:2018hin},  \cite{Gutperle:2020rty}, \cite{Gutperle:2018vdd}. It might be the case that some calculations are easier to perform using the language developed in this work. Also, finding string duals to theories that can be geometrically engineered, but do not have a nice expression in terms of an IR quiver field theory, appears as a very interesting goal. More generally, in the five-dimensional case we encounter at least two descriptions of the same system. Perhaps, it would be useful to develop an alternative language for the case of 3d ${\cal N}=4$ SCFTs along the lines of this work. Conversely, an alternative formulation in the case of half-maximal SCFTs in dimensions 1,2,4 along the lines of \cite{Assel:2011xz}, \cite{DHoker:2016ujz} might illuminate some aspects of them. It should be interesting to profit from these higher dimensional SCFTs for the construction of duals to more phenomenological QFTs, by adding a deformation in the string dual that induces interesting low energy dynamics.  Or similarly, to define new lower dimensional SCFTs by compactification, following for example \cite{Nunez:2001pt}. We hope to report on these issues in the near future.

\section*{Acknowledgments: }
The contents and presentation of this work much benefitted from extensive discussion with various colleagues. We are very happy to thank: Mohammad Akhond, Fabio Apruzzi, Lorenzo Coccia,  Yolanda Lozano, Niall Macpherson, Diego Rodr\'iguez-G\'omez, Sakura Sch\"afer-Nameki, Alessandro Tomasiello, Christoph Uhlemann, for their encouraging comments and sharing their knowledge with us.
We are supported by  STFC  grant  ST/T000813/1.

\appendix

\section{General comments on the backgrounds}\label{appendixcomments}

The AdS$_6$ background \eqref{background} is equivalent, up to an S-duality transformation, to the one presented in \cite{AGLMZ} (see also \cite{Niall-Tomasiello}). Differently to previous classifications of AdS$_6$ vacua (see \cite{Apruzzi:2014qva,DHoker:2016ujz}) the starting point of \cite{AGLMZ,Niall-Tomasiello} is a four-dimensional flat-space class with a round sphere on the internal space. The round sphere geometrically realises the SU(2) R-symmetry which is required by AdS$_d$ solutions with $\mathcal{N}=2$ and $d>4$, while the AdS$_6$ external space can be obtained by imposing Poincar\'e coordinates on the metric and that all the fluxes and warping function preserve the external-space isometries.
With these assumptions, all the Bianchi identities and the BPS equations reduces to a single partial differential equation, which is exactly \eqref{diffeq}.

In this section we show how to derive \eqref{background} from the AdS$_6$ solution presented in \cite{AGLMZ}, let's summarize that solution here. First of all, the supergravity equations of motion are invariant under the following rescaling of the metric and the fluxes:
\begin{equation}
\label{eq:rescaling_10D}
d s^2_{10} \to a^2 ds^2_{10} \, , \qquad  B_2 + i C_2 \to a^2 (B_2 + i C_2) \, ,
\end{equation}
and in particular we rescale the solution in \cite{AGLMZ} with  $a^2 = 3$. This is a convenient choice since it corresponds to $c_6=1$ in \cite{DHoker:2016ujz}, as we will see in the next section. The background therefore reads:
\begin{equation}
d s^2_{10} = e^{\frac{\Phi}{2}}  f_1(\sigma,\eta)\left[ds^2 (\text{AdS}_6) + f_2(\sigma,\eta)ds^2(S^2) + f_3(\sigma,\eta)(d\sigma^2+d\eta^2) \right]
\end{equation}
with
\begin{equation}
f_1 =2\sqrt{2}\sigma \left(\frac{3 \Lambda\partial_{\sigma}V}{\partial_{\eta}^2 V}\right)^{1/4} \, , \quad f_2 = \frac{\partial_{\sigma}V\partial_{\eta}^2V}{3\Lambda} \, , \quad f_3 = \frac{\partial_{\eta}^2V}{3\sigma\partial_{\sigma}V} \, ,
\end{equation}
while the fluxes are
\begin{align}
B_2 &= 4 \left(\frac{\sigma   \partial_{\sigma} V \left( \partial_{\eta} V \partial_{\sigma \eta}^2 V-3 \partial_{\eta}^2 V \partial_{\sigma} V \right)}{\Lambda}-V\right)\text{Vol}(S^2) , \quad C_2 = \frac{2}{9} \left(\eta -\frac{\sigma  \partial_{\sigma} V \partial_{\sigma \eta}^2 V}{\Lambda }\right) \text{Vol}(S^2), \nonumber \\[2mm]
C_0 &=-\frac{3 \partial_\sigma V \left(\partial_\eta V+\sigma  \partial_{\eta \sigma}^2 V\right)+\sigma  \partial_\eta V \partial_\eta^2 V}{18 \left(3 \partial_\sigma V \left(\sigma^2 (\partial_\eta^2 V)^2+\left(\partial_\eta V+\sigma  \partial_{\eta \sigma}^2 V\right)^2\right)+\sigma  \partial_\eta^2 V (\partial_\eta V)^2\right)} \, , \\[2mm]
e^{2 \Phi} &= \frac{108 \left(3 \partial_\sigma V \left(\sigma ^2 (\partial_\eta^2 V)^2+\left(\partial_\eta V+\sigma  \partial_{\eta \sigma}^2 V\right)^2\right)+\sigma  \partial_\eta^2 V (\partial_\eta V)^2\right)^2}{\sigma^2 \partial_\sigma V \partial_\eta^2 V \Lambda} \, . \nonumber 
\end{align}

Now, in order to get \eqref{background}, we need to perform an Sl$(2,\mathbb{R})$ transformation, which is parameterized by the matrix
\begin{equation}
S=\begin{bmatrix}
p & q \\
r & s
\end{bmatrix}
\, , \qquad ps- rq=1.\nonumber 
\end{equation}
The action of this duality keeps invariant the Einstein frame metric $d s^2_{10,E}=e^{-\frac{\Phi}{2}}d s^2_{10}$ and $F_5$, while it maps the other fields as following:
\begin{eqnarray}
& & \tau= C_0 + i e^{-\Phi}\longrightarrow \frac{p\tau +q}{r \tau + s},\\[2mm]
& & \begin{pmatrix}
C_{2}\\
B_{2}
\end{pmatrix} \longrightarrow
S  \begin{pmatrix}
C_{2}\\
B_{2}
\end{pmatrix} \, .\nonumber
\end{eqnarray}
It is immediate to notice that the Sl$(2,\mathbb{R})$ transformation we are looking for is a S-duality
\begin{equation}
S=\begin{bmatrix}
0 & -1 \\
1 & 0
\end{bmatrix}, \nonumber
\end{equation}
indeed, in terms of the \text{old} fields, the one in \eqref{background} are given by
\begin{eqnarray}
& & B_{2,\text{new}}= C_{2,\text{old}} \, , \qquad \quad \qquad \, \, \, \, \, \quad C_{2,\text{new}}=-B_{2,\text{old}} \, ,\\
& & C_{0,\text{new}}= -\frac{C_{0 , \text{old}}}{C_{0 , \text{old}}^2+ e^{-2 \Phi_{\text{old}}}} \, , \quad \quad  e^{-\Phi_{\text{new}}}=\frac{e^{-\Phi_{\text{old}}}}{ C_{0 , \text{old}}^2 + e^{-2\Phi_{\text{old}}}} \,.\nonumber
\end{eqnarray}

\section{Map to DGKU}
\label{sec:to_DGKU}

In this section we will show how it is possible to map the background in \eqref{background} to the DGKU solution \cite{DHoker:2016ujz}. Let's start by reviewing it.

\subsection{The DGKU solution}

The DGKU solution parameterize the Riemann surface in the internal space with a complex coordinate $w$, and it is entirely specified once two holomorphic functions $\mathcal{A}_\pm(w)$ are given. 
The metric
\begin{equation}
d s^2_{10} = e^{\frac{\Phi}{2}}  f_1(w,\bar{w}) \left[ds^2(\text{AdS}_6) + f_2(w,\bar{w}) d s^2 (S^2) + f_3(w,\bar{w}) d w d \bar{w} \right]
\end{equation}
is given by the following warping functions
\begin{equation}
\label{eq:DGKU_warpings}
f_1 = \frac{|\partial_w \mathcal{G}| \sqrt{1- R^2}}{\kappa \sqrt{R}}  , \quad f_2 = \frac{1}{9} \left(\frac{1-R}{1+R}\right)^2 \, , \quad f_3 = \frac{4 \kappa^4 R}{|\partial_w \mathcal{G}|^2 (1- R^2)^2}\,
\end{equation}
where
\begin{equation}
\label{eq:DGKU_functions}
\begin{split}
&\mathcal{G} = |\mathcal{A}_+|^2-|\mathcal{A}_-|^2+2 \text{Re} \mathcal{B} \, , \qquad \kappa^2 = - \partial_w \partial_{\bar{w}} \mathcal{G} = |\partial_w\mathcal{A}_-|^2-|\partial_w\mathcal{A}_+|^2 \, , \\
&\partial_w \mathcal{B} = \mathcal{A}_+\partial_w\mathcal{A}_- - \mathcal{A}_-\partial_w\mathcal{A}_+ \, , \qquad \, R+R^{-1} = 2 + \frac{6 \kappa^2 \mathcal{G} }{|\partial_w \mathcal{G}|^2} \, ,
\end{split}
\end{equation}
while the fluxes are given by:
\begin{equation}
\label{eq:DGKUfluxes}
\begin{split}
\tau= C_0 + i e^{-\Phi} =& -i \frac{\partial_w(\mathcal{A}_++\mathcal{A}_-) \partial_{\bar{w}} \mathcal{G} - R \partial_{\bar{w}}(\bar{\mathcal{A}}_++\bar{\mathcal{A}}_-) \partial_w \mathcal{G}}{\partial_w(\mathcal{A}_+-\mathcal{A}_-) \partial_{\bar{w}} \mathcal{G} + R \partial_{\bar{w}}(\bar{\mathcal{A}}_+-\bar{\mathcal{A}}_-) \partial_w \mathcal{G}} \, , \\
B_2 +i C_2 =& \frac{2}{3}i\left( \left(\frac{1-R}{1+R}\right)^2 \frac{\partial_w\mathcal{A}_+\partial_{\bar{w}} \mathcal{G}+\partial_{\bar{w}}\bar{\mathcal{A}}_-\partial_w \mathcal{G}}{3 \kappa^2}-\bar{\mathcal{A}}_--\mathcal{A}_+\right) \text{Vol}(S^2) \, .
\end{split}
\end{equation}
We have set the AdS$_6$ radius equal to one.
Notice that the entire solution is invariant under reparameterization if the complex coordinate $w \to z(w)$, which means that in principle we can use one of the holomorphic function (or a combination of them) as a definition of the complex coordinate.

\subsection{Matching the solutions}

In this section we will show how to match \eqref{background} with \eqref{eq:DGKU_warpings}-\eqref{eq:DGKUfluxes}. 
By equating the warping functions $f_1,f_2$ we get the following conditions
\begin{equation}
\label{eq:match1}
\mathcal{G} = 4\sigma^2 \partial_{\sigma} V \, ,
\qquad \frac{\kappa^2}{|\partial_w \mathcal{G}|^2} = \frac{1}{2 \sigma^2} \frac{\partial _{\eta}^2 V}{\Lambda-3\partial _{\eta}^2 V\partial _{\sigma }V} \, .
\end{equation}
Since the warping $f_3$ is actually coordinate dependent, we need to keep also the metric factor to make a comparison $f_3 d s^2 (\mathbb{C})$. Using the definition of $R$ we have that
\begin{equation}
\frac{2}{3} \frac{\kappa^2}{G} d w d \bar{w} = \frac{\partial_{\eta}^2V}{3\sigma\partial_{\sigma}V} (d \sigma^2 + d \eta^2) 
\end{equation}
and therefore, from equations $\eqref{eq:match1}$ we can write
\begin{equation}
|\partial_w \mathcal{G}|^2 d w d \bar{w} = \left( \left(\partial_\eta \mathcal{G} \right)^2+\left(\partial_\sigma \mathcal{G}\right)^2 \right) (d \sigma^2 + d \eta^2) \, .
\end{equation}
It is now immediate to notice that if we define a complex variable $z=\sigma - i \eta$ this consistency relation is automatically solved. Since the DGKU solution is defined up to a change of complex variables, we can consistently identify $w=z$ from now on.

Let's now consider the flux from \eqref{background}
\begin{equation}
B_2 + i C_2 = \frac{2}{3} i \left(6 V-\frac{i}{3} \eta- i \sigma   \partial_{\sigma} V\frac{ \left(-\frac{1}{3}- 6 i \partial_{\eta} V \right) \partial_{\sigma \eta}^2 V+18 i \partial_{\sigma} V \partial_{\eta}^2 V }{\Lambda}\right)\text{Vol}(S^2) \, .
\end{equation}
and compare it with \eqref{eq:DGKUfluxes} we have that the two expressions match if we set
\begin{equation}
\label{Apm_eq}
\mathcal{A}_++\bar{\mathcal{A}}_- = \frac{i}{3} \eta-6 \partial_\sigma (\sigma V) \, .
\end{equation}
Notice that $\sigma V = \hat{V}$ is the harmonic function defined in \eqref{eqfinal} and, since it is also real, it defines just one holomorphic function $\mathcal{V}(z)$
\begin{equation}
\sigma V = \mathcal{V}(z) + \overline{\mathcal{V}(z)}
\end{equation}
as already noticed in \eqref{potenciales}.
Using this condition and the fact that $\mathcal{A}_+$ is holomorphic while $\bar{\mathcal{A}}_-$ is anti-holomorphic, we have that \eqref{Apm_eq} completely defines $\mathcal{A}_\pm$ in terms of $\sigma V$ and the new coordinate $z$:
\begin{equation}
\label{eq:A1A2def}
\mathcal{A}_\pm = \mp \frac{z}{6} - 6\partial_z (\sigma V) \, .
\end{equation}
With these definitions one can check that the axion-dilaton expressions are identical.

Notice that $\partial_{\bar{z}}\mathcal{A}_\pm=0$ if $\partial_{\bar{z}} \partial_z \hat{V}=0$ and viceversa. In the solution we consider in the body of the text however we have that $\hat{V}$ is not harmonic everywhere, indeed using \eqref{identities1} we have
\begin{equation}
\label{eq:electrostatic_problem_2}
(\partial^2_\sigma + \partial^2_\eta) \hat{V} = \delta(\sigma) {\cal R}(\eta)
\end{equation}
which is due to the presence of the charge distribution at $\sigma=0$. This means that in $\sigma = 0$ the functions $\mathcal{A}_\pm$ are not holomorphic anymore; this is reflected in \eqref{potenciales}, where we have a different complex function depending on the sign of $\sigma$. In the next section we will see, however, that it is possible to extend the solutions in \cite{DHoker:2016ujz} across $\sigma=0$ in a natural way.

In order to compare the two backgrounds we had to impose $w=z$. As a consequence, we have that the two holomorphic functions $\mathcal{A}_\pm$ are defined just in term of one holomorphic function (i.e. $\mathcal{V}$) and the coordinate $z$. This means that if we perform a generic change of complex coordinate $z \to 6 \mathcal{F}(z)$ we get the general expression for \eqref{eq:A1A2def}:
\begin{equation}
\label{eq:A1A2gen}
\mathcal{A}_\pm = \mp \mathcal{F} - \frac{\partial_z (\sigma V)}{\partial_z \mathcal{F}} \, ,
\end{equation}
which defines two generic holomorphic functions in terms of two generic holomorphic functions.

\subsection{Examples}

In the DGKU formalism, the branes are given by the poles of the holomorphic functions $\mathcal{A}_\pm(w)$, which are all localized on the real axis of the Poincar\'e half-plane, and the value of the residue of $\partial_w \mathcal{A}(w)$ at the pole is the charge of the $(p,q)$-brane. In this section, we will see how solutions with three and four poles are mapped in terms of our potential $V$. We refer to \cite{Uhlemann:2019ypp} for the specific form of the solutions we are considering.

\subsubsection{$T_N$ theory}
\label{subsec:Tn_Uhul}

The $T_N$ theory is given by a three pole solutions
\begin{equation}
\mathcal{A}_\pm = \frac{3 N}{8 \pi } (\pm \log (w-1)+(\mp 1-i) \log (w+1)+i \log (2 w)) \, .
\end{equation}
where the poles are at $w=1,0,-1$. The coordinate $z = \sigma - i \eta$ is defined as
\begin{equation}
z = -3 (\mathcal{A}_+-\mathcal{A}_-) = \frac{9 N}{4 \pi} \log \left( \frac{1+w}{1-w} \right) \quad \Rightarrow \quad w = \coth \left(\frac{2 \pi  z}{9 N}\right) \, .
\end{equation}
Notice that the imaginary axis for $w$ becomes the interval $\eta \in (0, 9 N / 4)$ at $\sigma = 0$, while the real axis, which is the space-time boundary, is mapped as following
\begin{equation}
\begin{split}
w \in (-1,1) \quad &\Rightarrow \quad \eta = \frac{9}{4}N \, , \, \sigma \in (- \infty, \infty) \, , \\ \nonumber
w \in (-\infty,-1) \cup (1,\infty) \quad &\Rightarrow \quad \eta = 0 \, , \, \quad \sigma \in (- \infty, \infty) \, .
\end{split}
\end{equation}
So the space-time boundary in the $w$ coordinate is consistently mapped in the space-time boundary for $\sigma$ and $\eta$.

The potential is defined by the following equation
\begin{equation}
\partial_z (\sigma V) = -\frac{\mathcal{A}_- + \mathcal{A}_+}{12} = -\frac{i N}{16 \pi } \log \left(e^{-\frac{4 \pi z}{9 N}}+1\right)
\end{equation}
which can be integrated leading to
\begin{align}
V =& \frac{9 i N^2 }{64 \pi ^2 \sigma } \left(\text{Li}_2\left(-e^{-\frac{4 \pi  (|\sigma|+ i\eta )}{9 N}}\right)-\text{Li}_2\left(-e^{\frac{-4 \pi  (|\sigma|- i\eta )}{9 N}}\right)\right) \nonumber \\[2mm]
=& \frac{9 N^2 }{32 \pi^2 \sigma}\sum _{k=1}^{\infty }  \frac{(-1)^{k+1}}{k^2} \sin \left(\frac{4k \pi}{9 N} \eta \right) e^{-\frac{4k \pi}{9 N}|\sigma|} \label{eq:T_N-U} \, ;
\end{align}
notice that we set to zero the integration constant as required by the boundary conditions and we had to introduce the absolute value of $\sigma$ since the $\text{Li}_2(w)$ is well defined when $|w|<1$.
Notice that if we define $P = \frac{9 N}{4}$ we have that \eqref{eq:T_N-U} is exactly of the form \eqref{eq:fourier_vhat}, and we can easily identify the coefficient of the Fourier expansion:
\begin{equation}
a_k = \frac{ P^2 }{18 \pi^2} \frac{(-1)^{k+1}}{k^2} \, .
\end{equation}

We can go a bit further and in particular match this case with the example in section \ref{example1}. Indeed if we set $N=1$ in \eqref{eq:V_TN} and we take the large $P$ limit,  with the caveat that $\eta$ and $\sigma$ can be of order $P$, we get a potential which is proportional to \eqref{eq:T_N-U} up to some numerical factors. We are following the steps that lead from eq.(\ref{potentialgeneral})to eq.(\ref{aproximated}) for an approximate potential. This is not surprising, indeed the gauge group for the $T_N$ theory is exactly the one in \ref{example1} with a an SU(2) flavor instead of the first gauge group with  group. However, the contribution obtained by changing the quiver is subleading in the large $P$ limit and it can be ignored.

\subsubsection{$+_{MN}$ theory}
\label{subsec:+MN_Uh}

The $+_{MN}$ theory is a four-poles solution where the poles are at $w=0,\frac{2}{3},\frac12,1$. The holomorphic functions are given by
\begin{equation}
\mathcal{A}_\pm = \frac{3}{8 \pi } (\pm M(\log (3 w-2)-\log w)+ i N (\log (2 w-1)-\log (w-1)))
\end{equation} 
and, similarly to the previous example, we can define $z= \sigma - i \eta$ from them:
\begin{equation}
z = -3 (\mathcal{A}_+-\mathcal{A}_-)= \frac{9 M}{4 \pi }  (\log (w)-\log (3 w-2)) \quad \Rightarrow \quad w=\frac{2}{3} \left(\frac{1}{3 e^{\frac{4 \pi  z}{9 M}}-1}+1\right) \, .
\end{equation}
We also have
\begin{equation}
\partial_z (\sigma V) = -\frac{\mathcal{A}_- + \mathcal{A}_+}{12} = \frac{ i N \log \left(-\tanh \left(\frac{2 \pi z}{9 M}\right)\right)}{16 \pi }
\end{equation}
which can be integrated obtaining
\begin{align}
V =& \frac{9 i M N}{64 \pi ^2 \sigma } \left(\text{Li}_2\left(-e^{-\frac{4 \pi  (|\sigma| + i \eta )}{9 M}}\right)-\text{Li}_2\left(e^{-\frac{4 \pi  (|\sigma| + i \eta )}{9 M}}\right)-\text{Li}_2\left(-e^{-\frac{4 \pi  (|\sigma|- i \eta)}{9 M}}\right)+\text{Li}_2\left(e^{-\frac{4 \pi  (|\sigma|- i \eta)}{9 M}}\right)\right) \nonumber \\[2mm]
=& \frac{9 M N}{32 \pi^2 \sigma} \sum_{k=1}^\infty \frac{1-(-1)^k}{k^2} \sin \left( \frac{4 \pi k }{9 M} \eta \right) e^{-\frac{4 \pi k }{9 M} |\sigma|} \label{eq_V+mn_U}
\end{align}
where again we have set to zero the integration constant and regularized the Li$_2$. Again, this potential is exactly of the form \eqref{eq:fourier_vhat} once we identify $P = \frac{9 M}{4}$ and
\begin{equation}
a_k = \frac{ M P}{8 \pi ^2} \frac{1-(-1)^k}{k^2} \, .
\end{equation}

Following the steps that lead from eq.(\ref{potentialgeneral})to eq.(\ref{aproximated}) for an approximate potential, we
notice that the potential \eqref{eq_V+mn_U} matches exactly the one in \eqref{eq:+MN} in the large $P$ limit, up to an overall numerical factor.

\section{Some useful identities}\label{usefulidentities}
In this appendix we quote some useful calculations. We work with the two potentials defined  by eq.(\ref{change1}) and  find,
\begin{eqnarray}
& & \hat{V}= \sum_{k=1}^\infty  a_k \sin\left(\frac{k \pi\eta}{P}\right) e^{-\frac{k \pi |\sigma|}{P}},\label{identities1}\\
& & \partial_\sigma \hat{V}=-\sum_{k=1}^\infty  a_k\left(\frac{k \pi}{P}\right) sg(\sigma)\sin\left(\frac{k \pi\eta}{P}\right) e^{-\frac{k \pi |\sigma|}{P}},\nonumber\\
& & \partial_\eta \hat{V}=\sum_{k=1}^\infty  a_k\left(\frac{k \pi}{P}\right) \cos\left(\frac{k \pi\eta}{P}\right) e^{-\frac{k \pi |\sigma|}{P}},\nonumber\\
& & \partial_\sigma\partial_\eta \hat{V}= 
-\sum_{k=1}^\infty  a_k\left(\frac{k^2 \pi^2}{P^2} \right)sg(\sigma)\cos\left(\frac{k \pi\eta}{P}\right) e^{-\frac{k \pi |\sigma|}{P}},\nonumber\\
& & \partial_\eta^2 \hat{V}= -\sum_{k=1}^\infty  a_k\left(\frac{k^2 \pi^2}{P^2} \right) \sin\left(\frac{k \pi\eta}{P}\right) e^{-\frac{k \pi |\sigma|}{P}},\nonumber\\
& & \partial_\sigma^2 \hat{V}= \sum_{k=1}^\infty  a_k\left(\frac{k \pi}{P} \right)\left(\frac{k\pi}{P} -\delta(\sigma) \right) \sin\left(\frac{k \pi\eta}{P}\right) e^{-\frac{k \pi |\sigma|}{P}}.\nonumber
\end{eqnarray}
Also, the following are useful,
\begin{eqnarray}
& & \partial_\sigma V= \frac{\sigma \partial_\sigma \hat{V} -\hat{V}}{\sigma^2},\;\;\;\;\partial_\eta V=\frac{\partial_\eta \hat{V}}{\sigma},\;\;\;\partial_\eta^2 V= \frac{\partial_\eta^2 \hat{V}}{\sigma}\label{identities2}\\
& & \partial_\sigma\partial_\eta V=\frac{\sigma \partial_\eta\partial_\sigma \hat{V} -\partial_\eta \hat{V}}{\sigma^2},\;\;\; \partial_\sigma^2 V=\frac{2\hat{V} -2 \sigma \partial_\sigma \hat{V} +\sigma^2\partial_\sigma^2\hat{V}}{\sigma^3}.\nonumber
\end{eqnarray}
We use these results are used in the analysis of the behaviour of the field strengths and potentials as required in the Page charges, see Section \ref{pagesection}.
\subsection{The field $B_2$}
We will use the identities above (\ref{identities1})-(\ref{identities2}), to study the $B_2$ field. Ignoring the volume of the two-sphere $\text{Vol}(S^2)$, the expression is,
\begin{eqnarray}
& & \frac{9}{2} B_2= \eta -\frac{(\sigma \partial_\sigma V)( \partial_\sigma\partial_\eta V)}{\Lambda}=\eta-\frac{( \sigma^2 \partial_\sigma \hat{V} -\sigma \hat{V})( \sigma\partial_\sigma\partial_\eta \hat{V} -\partial_\eta \hat{V})}{\Lambda \sigma^4}.\nonumber\\
& & \sigma^4 \Lambda= \sigma \left[ (\sigma\partial_\sigma\partial_\eta \hat{V} -\partial_\eta^2\hat{V})^2 + (\partial_\eta^2\hat{V})( 3\sigma\partial_\sigma \hat{V} -3\hat{V} -\sigma^2\partial_\sigma^2\hat{V})\right]
\end{eqnarray}
Replacing the expansions in eqs.(\ref{identities1}) we find
\begin{eqnarray}
& & \sigma^4 \Lambda= \sigma \left( {\cal M}^2 + {\cal N} {\cal S}\right),\nonumber\\
& & {\cal M}=\sum_{k=1}^\infty a_k \left( \frac{k^2\pi^2}{P^2}\right)\left( \sin\left( \frac{k\pi \eta}{P}\right)  + |\sigma| \cos\left( \frac{k\pi \eta}{P}\right) \right) e^{-\frac{k \pi |\sigma|}{P}} ,\nonumber\\
& & {\cal N}= \sum_{k=1}^\infty a_k \left( \frac{k^2\pi^2}{P^2}\right)\sin\left( \frac{k\pi \eta}{P}\right) e^{-\frac{k \pi |\sigma|}{P}},\nonumber\\
& & {\cal S}=\sum_{k=1}^\infty a_k \sin\left( \frac{k\pi \eta}{P}\right) e^{-\frac{k \pi |\sigma|}{P}} \left( 3+\frac{3k\pi|\sigma|}{P} +\frac{k^2\pi^2\sigma^2}{P^2}\right) .\nonumber
\end{eqnarray}
The reader than check that for $\sigma\to\pm \infty$ we have 
\begin{equation}
\sigma \Lambda= a_1^2e^{-2\frac{\pi|\sigma|}{P}}\frac{\pi^4}{P^4}.\label{lambdaidentity}
\end{equation}
Putting all together, we have for $B_2$,
\begin{eqnarray}
& &\frac{9}{2} B_2-\eta= \frac{{\cal P}{\cal Q}}{\Lambda \sigma^4},\label{B2identity}\\
& &  {\cal P}= \sum_{k=1}^\infty a_k \sin\left( \frac{k\pi \eta}{P}\right) \sigma e^{-\frac{k \pi |\sigma|}{P}} \left( 1+\frac{k\pi|\sigma|}{P}\right),\nonumber\\
& & {\cal Q}= \sum_{k=1}^\infty a_k \cos\left( \frac{k\pi \eta}{P}\right) \left(\frac{k\pi}{P}\right) e^{-\frac{k \pi |\sigma|}{P}} \left( 1+\frac{k\pi|\sigma|}{P}\right).\nonumber
\end{eqnarray}
Evaluating the field in $\sigma\to\infty$ we find
\begin{equation}
B_2(\pm\infty,\eta)= f_4(\pm\infty,\eta)=\frac{2}{9}\left[\eta -\frac{P}{\pi} \sin\left( \frac{\pi \eta}{P}\right) \cos\left( \frac{\pi \eta}{P}\right) \right].\label{identityB2}
\end{equation}
\subsection{The field $C_0$}
We have the expression,
\begin{eqnarray}
& & \frac{C_0}{18}=
\partial_\eta V+\frac{\sigma \partial_\sigma\partial_\eta V}{1+ \frac{\sigma \partial_\eta^2 V}{3\partial_\sigma V}}= \sum_{k=1}^\infty a_k \cos\left( \frac{k\pi \eta}{P}\right) e^{-\frac{k \pi |\sigma|}{P}}\left[\frac{k\pi}{P\sigma}  -\frac{\frac{k\pi}{P\sigma} +\frac{k^2\pi^2}{P^2}sgn(\sigma)    }{1+ \frac{{\cal C}}{{\cal B}}}\right],\label{C0exp}\\
& & {\cal C}=\sigma^2  \sum_{k=1}^\infty a_k \left( \frac{k^2\pi^2}{P^2}\right)\sin\left( \frac{k\pi \eta}{P}\right) e^{-\frac{k \pi |\sigma|}{P}},\;\;\;{\cal B}= 3 
\sum_{k=1}^\infty a_k \sin\left( \frac{k\pi \eta}{P}\right) e^{-\frac{k \pi |\sigma|}{P}}\left(1+\frac{k\pi |\sigma|}{P}\right).\nonumber  
\end{eqnarray}
We find that at $\sigma=\epsilon$ for very small $\epsilon$---and using eq.(\ref{rankfunction}) $2\pi k a_k= - P c_k$ we have
\begin{equation}
C_0(0,\eta)=f_7(0,\eta)= 9 \sum_{k=1}^\infty c_k \left( \frac{k\pi}{P}\right)\cos\left( \frac{k\pi \eta}{P}\right) = 9\partial_\eta {\cal R}(\eta).\label{identityC0}
\end{equation}
\subsection{The combination $C_2- B_2 C_0$}
We study now the combination appearing when calculating the Page charge for D5 branes.
Ignoring the volumes of the two-sphere  we have,
\begin{align}
\frac{C_2- B_2 C_0}{4}
=&
V- \eta\partial_\eta V +\frac{\sigma\partial_\sigma V -\eta\sigma \partial_\eta\partial_\sigma V}{1+\frac{\sigma \partial_\eta^2 V}{3\partial_\sigma V}}. \label{c2b2c01} \\[2mm]
\frac{\sigma(C_2- B_2 C_0)}{4}=&\hat{V}-\eta\partial_\eta \hat{V}   +\left[\frac{\sigma\partial_\sigma \hat {V} -\eta\sigma \partial_\eta\partial_\sigma \hat{V} -\hat{V} +\eta\partial_\eta \hat{V}}{1+\frac{\sigma^2 \partial_\eta^2 \hat{V}}{3\sigma \partial_\sigma \hat{V} - 3\hat{V} }}\right]\nonumber\\[2mm]
=&\sum_{k=1}^\infty a_k\left[ \sin\left( \frac{k\pi \eta}{P}\right)  -\left( \frac{k\pi\eta}{P}\right)\cos\left( \frac{k\pi \eta}{P}\right)  \right]e^{-\frac{k \pi |\sigma|}{P}}\left(1- \frac{1+ \frac{k \pi |\sigma|}{P}}{1+\frac{{\cal C}}{ {\cal B}} } \right). \nonumber
\end{align}
The expressions for ${\cal C},{\cal B}$ have been defined in eq.(\ref{C0exp}).
We can now evaluate,
\begin{eqnarray}
& & (C_2- B_2 C_0)\Big]_{\sigma=\epsilon}^{\sigma=-\epsilon}= f_5- f_7 f_4\Big]_{\sigma=\epsilon}^{\sigma=-\epsilon}=8 \sum_{k=1}^\infty a_k \frac{k \pi}{P}\left[ \sin\left( \frac{k\pi \eta}{P}\right)  -\left( \frac{k\pi\eta}{P}\right)\cos\left( \frac{k\pi \eta}{P}\right)  \right].\nonumber
\end{eqnarray}
Using eq.(\ref{C0exp}) we find after the large gauge transformation,
\begin{eqnarray}
& &  (C_2- (B_2 +\Delta)C_0)\Big]_{\sigma=\epsilon}^{\sigma=-\epsilon}= f_5- f_7( f_4+\Delta) \Big]_{\sigma=\epsilon}^{\sigma=-\epsilon}=\nonumber\\
& & 8 \sum_{k=1}^\infty a_k \frac{k \pi}{P}\left[ \sin\left( \frac{k\pi \eta}{P}\right)  -\left( \frac{k\pi}{P}\right)\cos\left( \frac{k\pi \eta}{P}\right) (\eta-\frac{9\Delta}{2})  \right].
\label{c2b2c02}
\end{eqnarray}
\subsection{The background after rescaling}
After the rescaling in eq.(\ref{rescalingxx}), the full configuration reads
\begin{eqnarray}
& & ds_{10,st}^2= f_1(\sigma,\eta)\Big[ds^2(\text{AdS}_6) + f_2(\sigma,\eta)ds^2(S^2) + f_3(\sigma,\eta)(d\sigma^2+d\eta^2) \Big],\;\;e^{-2\Phi}=f_6(\sigma,\eta) , \nonumber\\[2mm]
& & B_2=f_4(\sigma,\eta) \text{Vol}(S^2),\;\;C_2= f_5(\sigma,\eta) \text{Vol}(S^2),\;\;\; C_0= f_7(\sigma,\eta), \label{backgroundrescaled}\\[2mm]
& & f_1= \frac{3 \pi}{2}\sqrt{\sigma^2 +\frac{3\sigma \partial_\sigma V}{\partial^2_\eta V}},\;\; f_2= \frac{\partial_\sigma V \partial^2_\eta V}{3\Lambda},\;\;f_3= \frac{\partial^2_\eta V}{3\sigma \partial_\sigma V},\;\;\Lambda=\sigma(\partial_\sigma\partial_\eta V)^2 + (\partial_\sigma V-\sigma \partial^2_\sigma V)  \partial^2_\eta V.\nonumber\\[2mm]
& & f_4=\frac{\pi}{2}\left(\eta -\frac{(\sigma \partial_\sigma V) (\partial_\sigma\partial_\eta V)}{\Lambda} \right),\;\;\;\; f_5=\frac{\pi}{2}\left( V- \frac{\sigma\partial_\sigma V}{\Lambda} (\partial_\eta V (\partial_\sigma \partial_\eta V) -3 (\partial^2_\eta V)(\partial_\sigma V)) \right),\nonumber\\[2mm]
& & f_6=12 \frac{\sigma^2 \partial_\sigma V \partial^2_\eta V}{(3 \partial_\sigma V +\sigma \partial^2_\eta V)^2}\Lambda,\;\;\;\; f_7=2\left( \partial_\eta V + \frac{(3\sigma \partial_\sigma V) (\partial_\sigma\partial_\eta V )}{3\partial_\sigma V +\sigma \partial^2_\eta V}  \right).\nonumber
\end{eqnarray}

\section{Generic expressions for the central charge and Potential}\label{centralchargeappendix}
Let us perform an analysis for the holographic central charge of a generic CFT
with rank function given by,
 \[ {\cal R}(\eta) = \begin{cases} 
          N_1 \eta & 0\leq \eta \leq 1 \\
          N_l+ (N_{l+1} - N_l)(\eta-l) & l \leq \eta\leq l+1,\;\;\; l:=1,...., P-2\\
          N_{P-1}(P-\eta) & (P-1)\leq \eta\leq P .
       \end{cases}
    \]
Using eq.(\ref{eq:fourier_vhat}), we compute
\begin{align}
&k \pi a_k=\int_{0}^P {\cal R}(\eta)\sin\left( \frac{k\pi \eta}{P}\right)=\label{akgeneric}\\
 & \frac{P}{k \pi}\left[ - N_1 \cos \left( \frac{k\pi}{P}\right) + N_{P-1}\cos \left( \frac{k\pi (P-1)}{P}\right) +\sum_{s=1}^{P-2} N_s \cos \left( \frac{s k\pi}{P}\right)  - N_{s+1}\cos \left( \frac{k(s+1)\pi}{P}\right)       \right] + \nonumber\\
& \frac{P^2}{k^2 \pi^2}\Big[ N_1 \sin \left( \frac{k\pi}{P}\right) + N_{P-1}\sin \left( \frac{k\pi (P-1)}{P}\right) +\sum_{s=1}^{P-2} (N_s - N_{s+1}) \left(\sin\left( \frac{k\pi s}{P}\right)  -\sin\left( \frac{k\pi (s+1)}{P}\right) \right)\Big] .\nonumber
\end{align}
By inspection one finds that the second line of eq.(\ref{akgeneric}) vanishes. The final value of the generic $a_k$ is,
\begin{eqnarray}
a_k=\frac{P^2}{\pi^3 k^3}\sum_{s=1}^{P-1} c_s \sin \left( \frac{k\pi s}{P}\right) ,\label{akgen2}
\end{eqnarray}
where one can explicitly compute
\begin{equation}
\label{c_s_def}
c_s = (2 N_s - N_{s-1}-N_{s+1}),
\end{equation}
is the rank of flavor group at that node, and $N_0=N_P=0$.
For the examples in section \ref{sec:QFT} we find:
\begin{enumerate}[start=1,label={(\bfseries Ex\arabic*):}]
	\item $c_s= N P \delta_{s,P-1}$, to compare with eq.(\ref{ak1}).
	\item $c_s= N  (\delta_{s,1}+\delta_{s,P-1})$, to compare with eq.(\ref{ak2})
	\item $c_s= (N- j K ) \delta_{s,1}+ j \delta_{s, K}$, to compare with eq.(\ref{ak3}).
\end{enumerate}

 Since the coefficients $c_s$ does not depend on $k$, we can compute the central charge from eq.(\ref{cholfinal}) in full generality:
\begin{eqnarray}
& & c_{hol}= -\frac{P^4}{8\pi^{10}} \sum_{k=1}^{\infty} \sum_{l=1}^{P-1} \sum_{s=1}^{P-1} c_l c_s \left(\frac{ e^{i\frac{k \pi (l+s)}{P}} + e^{-i\frac{k \pi (l+s)}{P}} - e^{i\frac{k \pi (l-s)}{P}} - e^{-i\frac{k \pi (l-s)}{P}}   }{k^5}  \right)=\label{centralgeneric}\\
& &  -\frac{P^4}{8\pi^{10}} \sum_{l=1}^{P-1} \sum_{s=1}^{P-1} c_l c_s\left(  \text{Li}_5( e^{i\frac{\pi (l+s)}{P}}) +\text{Li}_5( e^{-i\frac{\pi (l+s)}{P}}) - \text{Li}_5( e^{i\frac{\pi (l-s)}{P}}) - \text{Li}_5( e^{-i\frac{\pi (l-s)}{P}}) \right).\nonumber
\end{eqnarray}
The terms for which $l=s$ produce the $2\zeta(5)$ that we found in the case studies. Depending on the coefficients $c_l, c_s$ (conversely, depending on the quiver SCFT) the limit of large $P$ will take different expressions.

The reader can check that with the identifications for $c_s$ below eq.(\ref{akgen2}), the values of the central charge for Examples 1,2---see eqs.(\ref{centralexample1}) and (\ref{centralexample2}),  are easily  recovered.
Recovering the result for Example 3 in eq.(\ref{centralexample3}) is slightly subtler, as it requires the leading order expansions:
\begin{eqnarray}
& & 2\zeta(5) - \text{Li}_5(e^{2\pi i/N})- \text{Li}_5(e^{-2\pi i/N})\sim \frac{4\pi^2 \zeta(3)}{N^2}+O(1/N^3),\nonumber\\[2mm]
& & \text{Li}_5(e^{\frac{i \pi (l N+1)}{N}} ) + \text{Li}_5(e^{\frac{-i \pi (l N+1)}{N}})- \text{Li}_5(e^{\frac{i \pi (l N-1)}{N}})- \text{Li}_5(e^{\frac{-i \pi (l N-1)}{N }}) \nonumber\\
& &  \sim \frac{2\pi i}{N} \left( \text{Li}_4(e^{i \pi l}) - \text{Li}_4(e^{-i \pi l}) \right)+O(1/N^2).\nonumber
\end{eqnarray}
This can be interpreted as the fact that in the large $N$ limit the kinks that are at finite distance from the boundaries can be interpreted as squashed to it, indeed looking at the general expression for the Free Energy (3.17) in \cite{Uhlemann:2019ypp} the $\zeta(3)$ and Li$_4$ contributions are obtained from the value of the gauge node at the boundaries.\\

\underline{\bf General Potential $\hat{V}$}
\\
We can compute a generic expression for the potential function $\hat{V}(\sigma,\eta)$ as defined in eq.(\ref{eq:fourier_vhat}). Indeed, using this together with the generic expression in eq.(\ref{akgen2}) we find,
\begin{align}
\hat{V}=& \frac{P^2}{4\pi^3}\sum_{s=1}^{P-1}\sum_{k=1}^\infty \frac{c_s}{k^3}\left[W^k + \bar{W}^k -Z^k -\bar{Z}^k  \right],\label{potentialgeneral} \\
=& \frac{P^2}{2\pi^3}\sum_{s=1}^{P-1} c_s \text{Re} \left(\text{Li}_3(W)-\text{Li}_3(Z) \right) \nonumber \\
W&= e^{-\frac{\pi (|\sigma |+ i\eta- i s)}{P}},\;\;\;\; Z= e^{-\frac{\pi (|\sigma |+ i\eta+ i s)}{P} }. \nonumber
\end{align}
Using the definition of $c_s$ for each example it is immediate to check that the expressions in eqs.(\ref{eq:V_TN}),(\ref{eq:+MN}) and (\ref{eqVcompl}) are obtained by specialising eq.(\ref{potentialgeneral}).

To close this study for generic expressions of $c_{hol}$ and $\hat{V}$, let us discuss an approximate expression for $\hat{V}$. Coming back to the expression in eq.(\ref{eq:fourier_vhat}) we may be interested in considering the sum of modes up to a maximum value $k_{max}=\Lambda= l P$. We scale $P\to \infty$ and $l\to 0$ such that $\Lambda$ is finite and large. In this way the quotient $\frac{k\pi}{P}$ is always very close to zero, being its largest value $l\pi$. 

Consider now the potential $\hat{V}$ for the Example 1. Using that $\sin x\sim x$ for small values of $x$, we have
\begin{eqnarray}
& & \hat{V}_1\approx \sum_{k=1}^{\Lambda} (-1)^{k+1}\frac{NP^3}{k^3 \pi^3}\sin\left(\frac{k\pi}{P}\right) \sin\left( \frac{k\pi\eta}{P}\right) e^{-\frac{k\pi |\sigma|}{P}}\approx\label{aproximated}\\
& & \sum_{k=1}^{\Lambda} (-1)^{k+1}\frac{NP^2}{2 i k^2 \pi^2}\left[ e^{-\frac{k\pi}{P} (|\sigma | - i\eta)}  - e^{-\frac{k\pi}{P} (|\sigma | + i\eta)} \right]=\nonumber\\
& & \frac{i N P^2}{2\pi^2} \left[\text{Li}_2 (- e^{-\frac{\pi}{P} (|\sigma | - i\eta)} ) - \text{Li}_2 (- e^{-\frac{\pi}{P} (|\sigma | + i\eta)}) \right].\nonumber
\end{eqnarray}
This last expression should be compared with the expression \ref{eq:T_N-U}. Following the same procedure for the potential of the Example 2, we find
\begin{equation}
\hat{V}_2\approx \frac{i N P}{2\pi^2} \left[ 
\text{Li}_2 ( e^{-\frac{\pi}{P} (|\sigma | + i\eta)}  - \text{Li}_2 (- e^{-\frac{\pi}{P} (|\sigma | - i\eta)}) +\text{Li}_2 (- e^{-\frac{\pi}{P} (|\sigma | - i\eta)}  - \text{Li}_2 (- e^{-\frac{\pi}{P} (|\sigma | + i\eta)})  
\right].
\end{equation}
This expression should be compared with that in eq.(\ref{eq_V+mn_U}).

\end{document}